\documentclass{emulateapj}

\def\gsim{\mathrel{\rlap{\lower 4pt \hbox{\hskip 1pt $\sim$}}\raise 1pt
\hbox {$>$}}}
\def\lsim{\mathrel{\rlap{\lower 4pt \hbox{\hskip 1pt $\sim$}}\raise 1pt
\hbox {$<$}}}

\begin{document}
\title{Probing Shock Breakout and Progenitors of 
Stripped-Envelope Supernovae through Their Early Radio Emissions}

\author{
Keiichi Maeda\altaffilmark{1}
}

\altaffiltext{1}{Kavli Institute for the Physics and Mathematics of the 
Universe (Kavli-IPMU), Todai Institutes for Advanced Study (TODIAS), University of Tokyo
5-1-5 Kashiwanoha, Kashiwa, Chiba 277-8583, Japan; 
keiichi.maeda@ipmu.jp .}

\begin{abstract}
We study properties of early radio emission from stripped-envelope supernovae (those of type IIb/Ib/Ic). We suggest there is a sub-class of stripped-envelope supernovae in their radio properties, including optically well-studied type Ic supernovae (SNe Ic) 2002ap and 2007gr, showing a rapid rise to a radio peak within $\sim 10$ days reaching to a low luminosity (at least an order of magnitude fainter than a majority of SNe IIb/Ib/Ic). They show a decline after the peak shallower than others while the spectral index is similar. We show that all these properties are naturally explained if the circumstellar material (CSM) density is low and therefore the forward shock is expanding into the CSM without deceleration. Since the forward shock velocity in this situation, as estimated from the radio properties, still records the maximum velocity of the SN ejecta following the shock breakout, observing these SNe in radio wavelengths provides new diagnostics on natures of the breakout and progenitor which otherwise requires a quite rapid follow-up in other wavelengths. The inferred post-shock breakout velocities of SNe Ic 2002ap and 2007gr are sub-relativistic, $\sim 0.3c$. These are higher than inferred for SN II 1987A, in line with suggested compact progenitors. However, these are lower than expected for a Wolf-Rayet (WR) progenitor. It may reflect a still-unresolved nature of the progenitors just before the explosion, and we suggest that the WR progenitor envelopes might have been inflated which could quickly reduce the maximum ejecta velocity from the initial shock breakout velocity. 
\end{abstract}


\section{Introduction}

Core-collapse supernovae (CC-SNe) are explosions of a massive star with the zero-age main-sequence mass ($M_{\rm ms}$) exceeding $\sim 8 M_{\odot}$. Their observational appearance is diverse, controlled by natures of the progenitor star, explosion, and the circumstellar environment. These are probably connected mutually through some intrinsic controlling factors, including at least the progenitor mass, metallicity, and the binarity. Clarifying these mutual relations is a key issue in the study of SNe. In this respect, stripped-envelope SNe (SNe IIb/Ib/Ic) \citep{filippenko1997} have been a target of various studies: Being an explosion of a star which has lost all or most of the hydrogen envelope before the explosion \citep{nomoto1993}, their origins are likely a mixture of different evolutionary paths (i.e., a single massive star with $M_{\rm ms} \gsim 25 M_{\odot}$ or a binary evolution with $M_{\rm ms} \lsim 25 M_{\odot}$). They show diverse properties in the explosion energies \citep{nomoto2010}, with the most energetic ones linked to Gamma-Ray Bursts (GRBs) \citep[e.g.,][for a review]{woosley2006}. In this paper, we address the following three issues in their natures. (1) The nature of shock breakout, (2) progenitor structure and its relation to the shock breakout signal, and (3) radio emissions and the CSM environment. 

The shock breakout emission is the first electromagnetic signal from SNe \citep{falk1977,klein1978}. When the shock wave launched in the deepest part of the progenitor emerges from the stellar surface, an intensive UV/X-ray flash comes out of the shock wave. The typical spectral energy of the emission is predicted to be sensitive to the progenitor radius, and for a Wolf-Rayet (WR) progenitor it is mostly in X-rays. The shock breakout phenomenon was predicted theoretically more than a few decades ago, but so far the detection is rare \citep{soderberg2008,schavinski2008,gezari2008,modjaz2009}, first inferred from the ionization structure around SN 1987A \citep{lundqvist1996} -- it is essentially a very brief transient event. This is now an important target of time-domain astronomy, including future optical survey proposals to catch the shock breakout signals from high-z SNe \citep[e.g.,][]{tominaga2011}. However, such optical surveys are not optimized for SNe from a WR progenitor (i.e., a leading progenitor scenario for SNe Ib/c), since it is an X-ray event lasting only for $\sim R_{*}/c \sim 10$ seconds (where $R_{*}$ is the progenitor radius, and $c$ the speed of light). 

The final stage of the stellar evolution has not been clarified for stripped-envelope SNe. Estimating the radius of a progenitor star provides important insight here, highlighted by the direct detection of an unprecedented blue supergiant (BSG) progenitor with $R_{*} \sim 50 R_{\odot}${ for SN II 1987A \citep[see,][for a review]{arnett1989}. Without the direct progenitor detection so far \citep[see, ][for a recent review]{smartt2009}, the shock breakout signal and the subsequent early optical emission have been used to estimate the size of the progenitors of SNe Ib/c \citep{soderberg2008,chevalier2008,modjaz2009,rabinak2011}. SN Ib 2008D is the only example where the shock breakout X-ray was convincingly detected \citep{soderberg2008,modjaz2009}, and the radius estimated here is still in debate \citep{chevalier2008, rabinak2011}. Any other independent information on the still-unresolved natures of the progenitors is highly valuable. 

Radio emissions have been detected from a number of nearby SNe IIb/Ib/Ic \citep[see,][for a review]{chevalier2006b}. The radio emission is created through the SN-CSM interaction, thus it is a strong tool to study the CSM environment. A relation between the time of the peak ($t_{\rm p}$) and the luminosity there ($L_{\rm p}$) can be used to estimate the size of the emitting region (thus the velocity of the shock wave at the SN-CSM interaction), and the CSM density \citep{chevalier1998}. Generally SNe of different spectral sub-classes occupy different regions in the $t_{\rm p} - L_{\rm p}$ plot, and SNe Ib/c (plus a part of SNe IIb, i.e., `compact' SNe IIb: `cIIb' hereafter) are characterized by relatively low $t_{\rm p}$ and large $L_{\rm p}$ for given $t_{\rm p}$ \citep{chevalier2006b,chevalier2010}. However, radio emission from SNe IIb/Ib/Ic is diverse in a sense that even excluding outliers they span more than an order of magnitude both in $t_{\rm p}$ and $L_{\rm p}$ ($t_{\rm p} \sim 10 - 100$ days, $L_{\rm p} \sim 10^{26} - 10^{28}$ erg s$^{-1}$ Hz$^{-1}$). Among SNe IIb/Ib/Ic, SN Ic 2002ap is a suggested outlier characterized by small $t_{\rm p}$ and $L_{\rm p}$. As we show in \S 2, SN Ic 2007gr shares similar properties with SN 2002ap, and these SNe show different features than others not only in $t_{\rm p}$ and $L_{\rm p}$. 

In this paper, we propose a new idea to explain the peculiarities of radio emission from SNe Ib/c with low radio luminosities and a short rising time (i.e., SN 2002ap, 2007gr), and link our interpretation with the properties of shock breakout and progenitor. In \S 2, we summarize radio properties of SNe 2002ap and 2007gr, suggesting that these two belong to a sub class of stripped-envelope SNe (in terms of radio properties). In \S 3, we explore consequences of the hydrodynamic interaction between the SN ejecta and CSM in the early phase, and show that the evolution there can be different for SNe exploding in low density CSM environment (SNe 2002ap and 2007gr) and high density CSM (other SNe IIb/Ib/c). In \S 4, we calculate multi-band radio light curves for SNe 2002ap and 2007gr, showing that their radio properties are naturally explained in terms of the hydrodynamic evolution discussed in \S 3. In \S 5, we discuss implications of our findings for natures of shock breakout and progenitor. The paper is closed in \S 6 with conclusions and discussion. Our scenario requires (relatively) inefficient acceleration of electrons at a shock, and discussion on this issue is given in Appendix. 

\section{Radio Properties of SNe 2002ap and 2007gr}

Properties of radio emissions from SNe are characterized by a relation between the peak luminosity and the peak date (normalized at $5$ GHz) \citep{chevalier1998,chevalier2006a}. This is especially useful diagnostics for SNe Ib/c \citep{chevalier2006b,chevalier2010}: Since the synchrotron self-absorption (SSA) is a dominant absorption process for these SNe exploding within a relatively low density CSM environment, a place in this $t_{\rm p} - L_{\rm p}$ plot provides an estimate of the size of the emitting region (thus the velocity of the shock wave). The velocity of the shock wave estimated this way is typically $\sim 0.1 c$ for SNe cIIb/Ib/Ic. At the same time, within the framework of the SSA, $t_{\rm p}$ is smaller for less dense CSM. 

SN 2002ap is a classical example of broad-line SN Ic in optical wavelengths, suggested to be more energetic than canonical SNe \citep{mazzali2002}. SN 2002ap is peculiar also in properties of radio emissions as compared to other SNe Ib/c \citep{berger2002,bjornsson2004}. While $L_{\rm p} \sim 10^{27}$ erg s$^{-1}$ Hz$^{-1}$ and $t_{\rm p} (\nu_{\rm p}/5 {\rm GHz}) \sim 30$ days for a majority of SNe Ib/c, in SN 2002ap $L_{\rm p} \sim 10^{25}$ erg s$^{-1}$ Hz$^{-1}$ and $t_{\rm p} (\nu_{\rm p}/5 {\rm GHz}) \sim 3$ days characterized by an extremely low luminosity and a fast rise to the peak. According to the SSA scaling relation, the CSM density is estimated to be lower than other SNe Ib/c at least by an order of magnitude. In the optically thin phase, a majority of SNe Ib/c show the radio spectral slope $\alpha \sim -1$ and the temporal slope $-1.5 \lsim \beta \lsim -1.3$ (where $L_{\nu} \propto \nu^{\alpha} t^{\beta}$) \citep{chevalier2006b}. However, SN 2002ap showed a shallow decay with $\beta \sim -0.9$ while the spectral index was similar to the others ($\alpha \sim -0.9$) \citep{berger2002} \citep[see also][]{bjornsson2004,chevalier2006b}. This peculiarity refuses a standard interpretation for SN 2002ap. There is only one theoretical interpretation suggested so far -- a slope of the relativistic electrons' energy distribution might have been different and flatter than in other SNe Ib/c \citep{bjornsson2004} \citep[see also][]{chevalier2006b}. 

We point out that SN Ic 2007gr shares the similar properties in radio with SN 2002ap, despite its optical properties belonging to a `normal' (or non broad-line) class \citep{hunter2008,valenti2008}. In Figure 1, we compare the multi-band radio light curves of SNe 2002ap \citep{berger2002} and 2007gr \citep{soderberg2010}. SN 2007gr showed $L_{\rm p} \sim 10^{26}$ erg s$^{-1}$ Hz$^{-1}$ and $t_{\rm p} (\nu_{\rm p}/5 {\rm GHz}) \sim 5$ days, both about an order of magnitude smaller than typically found for SNe Ib/c, placing this SN close to SN 2002ap in these properties. Furthermore, the radio spectral index and decay slope in optically-thin phases are similar to those of SN 2002ap. Not only in $t_{\rm p}$ and $L_{\rm p}$, but the temporal slopes are different in SNe 2002ap and 2007gr ($\beta \gsim  -1$) from other SNe Ib/c ($\beta \lsim -1.3$), while the spectral indexes are similar ($\alpha \sim -1$). 

The previous theoretical interpretation of the radio emission from SN 2002ap was based on the slow decay rate as mentioned above \citep{bjornsson2004}, and this is our motivation to investigate an alternative explanation for this behavior. However, it should be noted that the quality of the radio data of SN 2002ap, as well as that of SN 2007gr, does not allow very accurate determination of the decay rate. Fitting the radio light curve of SN 2002ap \citep{berger2002} during 4 - 50 days (8 points) by a function $f_{\nu} \propto t^{\beta}$, we obtain $\beta = -0.87 \pm 0.17$ at 8.46 GHz.\footnote[1]{To ensure the optically thin nature of the emission, we compute the decay rate in the highest frequency band data available.} The error indicated here is $1\sigma$. For SN 2007gr \citep{soderberg2010}, we obtain $\beta = -0.85 \pm 0.12$ (4 - 18 days, 5 points). For comparison, `radio-normal' SN cIIb 2011dh \citep{soderberg2012,krauss2012} shows $\beta = -1.13 \pm 0.16$ at 29.0 GHz (30 - 100 days, 4 points). So, the preferred value of the decay rate ($\beta$) is flatter by $\sim 0.3$ in SNe 2002ap and 2007gr than the canonical case. The deviation of SN 2002ap in the decay rate from the canonical value \citep[$\beta = - 1.3: $][]{chevalier2006b} is therefore 2.5$\sigma$ level, and for SN 2007gr it is above $3\sigma$. Given the small number of the data points, however, this nominal significance should be regarded as merely indicative. In any case, in this paper we will investigate the implications provided by this slow decay. 

As summarized in this section, SNe 2002ap and 2007gr (seem to) share the common properties, and we suggest they form a sub class of stripped-envelope SNe in the radio properties. Observationally this class of objects are rare (about 10\% of radio-detected stripped-envelope SNe), but they may well occupy intrinsically a large fraction of stripped-envelope SNe: SNe 2002ap and 2007gr are among the nearest examples of stripped-envelope SNe, and they would not have been detectable at $\sim 30$Mpc, i.e., the typical distance of most radio-detected SNe \citep{soderberg2010}. Future observations of nearby SNe in radio wavelengths will be critical to establish if the shallow decay is a common property of these radio low-luminosity stripped-envelope SNe. 

\section{Hydrodynamics in the Early Phase}

One of the striking features of radio emission from SNe 2002ap and 2007gr is the short rise time with the time scale less than 10 days. In this section we discuss the hydrodynamic evolution of the shocked region in this early phase. In previous studies, the self-similar solution for the interaction between the expanding SN ejecta and CSM has been assumed \citep{chevalier1982}. However, the basic assumption in the formalism would not apply in the very early phase of the expansion and/or low CSM density. The effect of the interaction on the hydrodynamic evolution becomes important only after a sufficient amount of CSM is swept up by the forward shock, and then the ejecta start being decelerated, following the self-similar solution which is eventually established. Before this phase, the ejecta feel the CSM almost as a vacuum, thus the ejecta are in the free expansion phase.\footnote[2]{The decelerated ejecta following the self-similar solution is frequently referred as the `free expansion' in many literatures since the deceleration is not large in the self-similar solution considered here (i.e., $R \propto t^{0.9}$) as compared to the Sedov phase (i.e., $R \propto t^{0.4}$ in CSM with constant density). In this paper, we refer the `truly' free expansion phase (i.e., $R \propto t$) before the establishment of the self-similar solution as the `free expansion' phase.} 

The density structure at the outermost ejecta can be approximated by 
\begin{eqnarray}
\rho_{\rm SN} & \sim & 8.3 \times 10^{-18} E_{51}^{3.59} 
\left(\frac{M_{\rm SN}}{M_{\odot}}\right)^{-2.59} 
\left(\frac{V}{0.3 c}\right)^{-10.18}
t_{\rm d}^{-3} \nonumber\\
& & \ {\rm g} \  \rm{cm}^{-3} \ , 
\end{eqnarray}
where $\rho_{SN}$ is the density of the SN ejecta at the velocity $V$, $E_{51}$ is the kinetic energy of the SN ejecta (in unit of $10^{51}$ erg), $M_{\rm SN}$ is the ejecta mass, and $t_{\rm d}$ is the time since the explosion in day \citep{matzner1999,chevalier2008}. 

The maximum velocity of the SN ejecta is determined by the shock breakout set by the radiative losses due to the shock breakout flash, unless there is any process that could alter the dynamics at the highest velocity ejecta after the breakout. Although detailed radiation hydrodynamic modeling is required to obtain the exact value and the result depends on details in the treatment of the physics involved, the first-order estimate can be obtained by analytic considerations \citep{matzner1999}: 
\begin{eqnarray}
V_{\rm SN} & \sim & 0.48 c  \left(\frac{\kappa}{0.34 {\rm cm}^2 {\rm g}^{-1}}\right)^{0.16} E_{51}^{0.58} 
\left(\frac{M_{\rm SN}}{M_{\odot}}\right)^{-0.42} \nonumber\\
& & \times \left(\frac{R_{\rm *}}{10 R_{\odot}}\right)^{-0.32} \ ,
\end{eqnarray}
where $R_{\rm *}$ is the radius of the progenitor (for that with a radiative envelope). 

Let us assume that the density distribution of the CSM is expressed by the steady wind solution, $\rho_{\rm CSM} = \dot M / 4 \pi r^{2} v_{\rm w}$ (where $\dot M$ and $v_{\rm w}$ are the mass loss rate and the wind velocity). Thus, 
\begin{equation}
\rho_{\rm CSM} = 5 \times 10^{11} A_{\rm *} r^{-2} \ {\rm g} \ {\rm cm}^{-3} \ ,
\end{equation} 
where $A_{\rm *} \sim 1$ is a reference value corresponding to typical Wolf-Rayet mass loss properties (i.e., $\dot M \sim 10^{-5} M_{\odot}$ yr$^{-1}$ and $v_{\rm w} \sim 1000$ km s$^{-1}$). 
Once the self-similar solution is established, the evolution of the velocity at the contact discontinuity follows the following form \citep{chevalier1982,chevalier2006b}: 
\begin{equation}
V_{\rm c} \sim 8 \times 10^9 E_{51}^{0.43} \left(\frac{M_{\rm SN}}{M_{\odot}}\right)^{-0.32} 
A_{\rm *}^{-0.12} t_{\rm d}^{-0.12} \ {\rm cm} \ {\rm s}^{-1} \ . 
\end{equation}

Since the self-similar solution describes the deceleration of the ejecta, the velocity here should be smaller than $V_{\rm SN}$. Thus, the self-similar solution does not apply if 
\begin{equation}
t \lsim t_{\rm dec} = 0.4 E_{51}^{3.58} \left(\frac{M_{\rm SN}}{M_{\odot}}\right)^{-2.67} A_{\rm *}^{-1} 
\left(\frac{V_{\rm SN}}{0.3 c}\right)^{-8.33} \ {\rm day} \ .
\end{equation}

Assuming $V_{SN} \gg 0.1 c$ and the typical value of a few $M_{\odot}$ for the ejecta mass and $A_{\rm *} \sim 1$, the strong interaction starts at latest a few days after the explosion, and thus the free expansion phase is negligible. This justifies the use of the self-similar solution for most of SNe. However, if $V_{\rm SN}\le 0.1c$, then even a typical WR wind density is not enough to decelerate the ejecta to reach to the self-similar solution within the time scale of $\sim 100$ days. Furthermore, a low CSM density prevents the SN ejecta being decelerated, and with $A_{*} \sim 0.01$ as inferred for sub class of SNe Ib/c (for example, the scaling relation of the SSA peaks implies that $A_{*} \sim 0.04$ for SN 2002ap), the ejecta do not experience significant deceleration for $\sim 100$ days even with $V_{\rm SN} \sim 0.2 c$.  
 
To confirm the above arguments, we have performed a series of hydrodynamic simulations for the early phase of the interaction. The ejecta structure is assumed to follow a power law up to the maximum ejecta velocity $V_{\rm SN}$ (equation 1), with $E_{51} = 1$ and $M_{\rm ej} = 3 M_{\odot}$. The CSM density distribution is assumed to be a power law as described by equation 3. We have varied $V_{\rm SN}$ and $A_{*}$ as parameters, choosing $V_{\rm SN} = 0.1c, 0.2c, 0.3c$ and $A_{*} = 0.01, 1, 100$ (thus 9 models were investigated in total). The interaction starts at 1 day since the explosion. Adiabatic Euler equations under spherical symmetry have been solved with the adiabatic index of 5/3, using the HLLC (Harten-Lax-van Leer Contact) solver \citep{toro1999} to treat the discontinuity and the shock waves \citep{maeda2002}. The number of meshes is 8500 linearly covering up to $7.8 \times 10^{16}$ cm, so the spatial resolution is about $10^{13}$ cm which is sufficient to resolve detailed structure of the interaction region throughout the computation. 

Figures 2 and 3 show the evolution of the forward shock velocity and radius as a function of time. It is seen that the above analytical estimate roughly explains the behavior of the shock wave velocity: Models with low $V_{\rm SN}$ and/or low $A_{*}$ do not follow the self-similar deceleration, but rather show an almost constant forward shock velocity. According to equation 5, we expect $t_{\rm dec}$ as follows: For $A_{*} = 100$, $t_{\rm dec} < 2 $ days for all values of $V_{\rm SN}$, for $A_{*} = 1$, $t_{\rm dec} < 1$ day for $V_{\rm SN} > 0.2 c$ but $\sim 200$ days for $V_{\rm SN} = 0.1c$, and for $A_{*} = 0.01$, $t_{\rm dec} \sim 2, 60, 20000$ days for $V_{\rm SN} = 0.3c, 0.2c, 0.1c$, respectively. In the simulations, for $A_{*} = 100$ the self-similar solution is approximately reached for all the cases. For $A_{*} = 1$ the high velocity models with $V_{\rm SN} \gsim 0.2c$ quickly follow the self-similar solution but the model with $V_{\rm SN} = 0.1c$ starts to be decelerated at $t \sim 100$ days. For $A_{*} = 0.01$ the model with $V_{\rm SN} = 0.1c$ never experiences the deceleration during the simulated period of time, and the model with $V_{\rm SN} = 0.2c$ does not show a sign of deceleration before $t \sim 40$ days. 

Figure 4 shows the evolution of the velocity profiles for SN models expanding into the low density CSM with $A_{*} = 0.01$, with the SN ejecta velocity $V_{\rm SN} = 0.1c, 0.2c$, and $0.3c$. One immediately notices that all these three models do not follow the self-similar solution in the early phase, since in such a case there must be no difference according to different $V_{\rm SN}$. In our simulations, the velocity of the leading edge of the ejecta is not identical to the initial value (i.e., $V_{\rm SN}$). At 6 days after the explosion (which may depend on the initial setup), the forward shock begins to develop, and the maximum velocity of the ejecta is $\sim 75,000$ km s$^{-1}$ for $V_{\rm SN} = 0.3 c$, $65,000$ km s$^{-1}$ for $V_{\rm SN} = 0.2c$, and $45,000$ km s$^{-1}$ for $V_{\rm SN} = 0.1 c$. Thus, the kinetic energy is redistributed in a way that it is decelerated for a large value of $V_{\rm SN}$ and accelerated for a small value of $V_{\rm SN}$. This kinetic energy redistribution may partly stem from a specific simulation setup, and details of this process could be dependent on the initial conditions. Anyway, once the velocity profile is quickly adjusted, there is no further significant evolution here, and following arguments should not be sensitive to details in the numerical setup. The forward shock is formed with the velocity roughly $80 \%$ of the (redistributed) maximum velocity of the ejecta, and the forward shock expands into the CSM at almost constant velocity without deceleration. 

It has been argued that the free-expansion phase is negligible for SNe Ib/c, with the shock breakout velocity $V_{\rm SN} > 0.3 c$ estimated with equation 2 or similar expressions, under the assumption $R_{*} \sim R_{\odot}$ \citep{chevalier2006b}. However, it has not been clarified yet to what extent the analytic expression on the shock breakout physics is accurate: the expression is an order-of-magnitude estimate with various assumptions. The progenitor structure just before the explosion is in debate, without direct detection of the SNe Ib/c progenitors so far. For example, a peculiar SN Ib 2006jc showed a luminous blue variable (LBV)-like eruption about two years before the SN explosion \citep{pastorello2007}. A mechanism of such an outburst of the proposed WCO Wolf-Rayet progenitor \citep{tominaga2008} has not been clarified yet, highlighting our ignorance of the final evolution of WR stars toward the SN explosion. 

Considering all these uncertainties in the shock breakout dynamics and progenitor scenarios, our motivation in this paper is to tackle to these issues from the opposite direction: We first provide possible, observationally-based information on these issues, then compare this with the theoretical expectations. Our strategy is the following. Rather than starting with the shock velocity based on equation 2 (or similar expressions), we first compare the velocity of the forward shock estimated from the radio data and that predicted by the self-similar decelerating solution. The two should agree if the dynamics is in the self-similar phase, while the radio-derived velocity will be lower than the self-similar expectation if the ejecta are in the free expansion phase. We then examine the radio spectrum and light curve properties of these SNe to see if the free-expansion solution provides a consistent view. Once we could obtain information on the free-expansion velocity through these analyses, then we could translate this information to the velocity of the shock breakout. This could then be used to calibrate the shock breakout physics (e.g., equation 2) and/or natures of progenitors. 

Useful diagnostics of the expansion of the forward shock, especially for stripped-envelope SNe, is provided by the observed peak date -- peak radio luminosity relation. Figure 5 is a reproduction of the diagram from \citet{soderberg2012} \citep[see also][]{chevalier1998,chevalier2006a}. Over plotted lines in the figure are the estimates which divide the (truly) free expansion phase and the self-similar deceleration phase using equation 5, for different choices of the ejecta mass. Below these lines, the observationally derived velocity (through the SSA model) is smaller than the self-similar prediction, thus SNe in this range are very likely in the free-expansion phase.

Figure 5 shows that SNe Ib/c as well as cIIb generally follow the self-similar solution with $E_{51} \sim 1$ and $M_{\rm ej} \sim 1 - 3 M_{\odot}$. An exception is SN 2002ap (as well as 1987A) which falls well below the self-similar prediction. Since different SNe have different properties in $E_{51}$ and $M_{\rm ej}$, one has to compare the self-similar evolution and the radio properties on case by case basis. With $E_{51} \sim 5$ and $M_{\rm ej} \sim 3 M_{\odot}$ for SN 2002ap estimated through the optical modeling \citep{mazzali2002}, the discrepancy between the self-similar solution and the observed radio properties is even larger. The same argument applies to SNe 2007gr ($E_{51} \sim 2$, $M_{\rm ej} \sim 2 M_{\odot}$) \citep{hunter2008,valenti2008} and 1987A ($E_{51} \sim 1.4$, $M_{\rm ej} \sim 14 M_{\odot}$) \citep[e.g.,][]{blinnikov2000,maeda2002}, which showed that their radio properties are also below the self-similar case. Thus, we suggest that the shock wave evolution of SNe 2002ap and 2007gr (and 1987A) were in the free-expansion phase in these early epochs responsible for the early radio emission covering the SSA peak.

\section{Radio emission from SNe within low density CSM}

Results from previous studies on radio properties of SN 2002ap have been controversial. In the following, we show that a self-consistent picture can be obtained by considering the `free expansion' which was not taken into account in the previous studies. We show further that the same solution could apply to SN 2007gr as well. In our scenario, the peculiar features of SNe 2002ap and 2007gr are natural consequences of SNe exploding within relatively low density CSM. 

We calculate radio emissions from SNe Ib/c as follows. The basic formalisms have been developed by \citet{fransson1998} and \citet{bjornsson2004}, and specific prescriptions used here are given by \citet{maeda2012} \citep[see also,][]{chevalier1998,soderberg2005,chevalier2006a,chevalier2006b,soderberg2010}. The synchrotron radio luminosity $\nu L_{\nu}$ in the optically thin phase is given as 
\begin{equation}
\nu L_{\nu} \sim \pi R_{\rm sh}^2 V_{\rm sh} n_{\rm rel} \gamma_{\nu}^{2-p} m_{\rm e} c^2 
\left[1 + \frac{t_{\rm synch (\nu)}}{t} + 
\frac{t_{\rm synch (\nu)}}{t_{\rm other} (\nu)}\right]^{-1} \ .
\end{equation}
Here $R_{\rm sh}$ and $V_{\rm sh}$ are the position and the velocity of the forward shock, $n_{\rm rel}$ is the number density of the relativistic electrons. The relativistic electrons are assumed to follow a power law distribution with the index $p$ as a function of the energy (note that throughout this paper we use $p$ as the intrinsic power law index, before being altered by cooling effects). $t_{\rm synch}$ is the synchrotron cooling time scale. $t_{\rm other}$ is the time scale for other energy loss processes which do not emit at the radio frequency (i.e., the IC scattering under the situation investigated here). 
The Lorentz factor of the electrons emitting at frequency $\nu$ is 
$\gamma_{\nu} \sim 80 \nu_{10}^{0.5} B^{-0.5}$ (here $\nu_{10} = \nu/10^{10}$ Hz and B is in gauss). 

In SNe IIb/Ib/Ic, the assumption of equipartition has been proven to work well \citep{fransson1998,soderberg2005,chevalier2006b}. The energy densities of the relativistic electrons and the magnetic field behind the shock wave are proportional to the thermal energy density created by the shock wave, while the proportional coefficients ($\epsilon_{\rm e}$ and $\epsilon_{B}$) are generally found to be lower than the full equipartition: Typically $\epsilon_{\rm e} = \epsilon_{B} = 0.1$ is assumed for simplicity, while \cite{fransson1998} and \citet{maeda2012} have suggested $\epsilon_{\rm e} \lsim 0.01$ and $\epsilon_{B} \sim 0.01 - 0.15$. Our general arguments are independent from the values of $\epsilon_{\rm e}$ and $\epsilon_{B}$, for which further discussion is given in Appendix. The amplified magnetic field strength and the relativistic electron density are expressed as follows: 
\begin{eqnarray}
B & \sim & 2.2 \times 10^{6} \epsilon_{B, -1}^{0.5} A_{\rm *}^{0.5} 
\frac{V_{\rm sh}}{R_{\rm sh}} {\rm gauss}\ , \\
n_{\rm rel} & \sim & 2.4 \times 10^{17} \frac{p-2}{p-1} \epsilon_{e, -1} A_{\rm *} 
\left(\frac{V_{\rm sh}}{R_{\rm sh}}\right)^2 {\rm cm}^{-3} \ .
\end{eqnarray}

Substituting these expressions into equation 6, the synchrotron emission is scaled as 
\begin{equation}
L_{\nu} \propto V_{\rm sh}^3 \nu^{-p/2} B^{(p-2)/2} 
\left[1 + \frac{t_{\rm synch (\nu)}}{t} 
+ \frac{t_{\rm synch (\nu)}}{t_{\rm other} (\nu)}\right]^{-1} \ .
\end{equation}
Note that so far no assumption has been made on the time-dependence of the expansion of the interaction region. 

In these SNe, the main cooling agencies are the synchrotron cooling and the inverse Compton (IC) scattering. These cooling time scales are estimated as 
\begin{eqnarray}
t_{\rm synch} (\nu) & \sim & 110 \nu_{10}^{-0.5} B^{-1.5} \ {\rm days} \ ,\\
t_{\rm IC} (\nu) & \sim & 1.7 \nu_{10}^{-0.5} B^{0.5} 
\left(\frac{L_{\rm SN}}{10^{42} {\rm erg s}^{-1}}\right)^{-1} \nonumber\\
& & \times \left(\frac{R_{\rm sh}}{10^{15} {\rm cm}}\right)^{2} {\rm days} \ .
\end{eqnarray}

Now, the spectral index ($\alpha$) and the temporal slope ($\beta$) of the synchrotron emission can be computed (here $\L_{\nu} \propto \nu^{\alpha} t^{\beta}$). Describing  the expansion dynamics as $R_{\rm sh} \propto t^{m}$ (therefore $V_{\rm sh} \propto t^{m-1}$), the result is summarized in Table 1. 

A majority of SNe Ib/c show the spectral index $\alpha \sim -1$ and the temporal slope $-1.5 \lsim \beta \lsim - 1.3$ \citep{chevalier2006b}, being consistent with the adiabatic expansion following the self-similar deceleration ($m \sim 0.9$) with $p \sim 3$. 
The `low-density CSM' SNe Ic, 2002ap and 2007gr, differ than the other cases in the temporal slope: they show a rather shallow decay, $\beta \sim -1$, while the spectral index is similar to the others ($\alpha \sim -1$). One solution is to assume that the intrinsic slope of the relativistic electron distribution is flatter ($p \sim 2$) than other SNe Ib/c ($p \sim 3$) and that the electrons emitting in the radio frequency are in the IC cooling regime \citep{bjornsson2004}, while assuming the self-similar dynamics ($m \sim 0.9$). In this case, the spectral slope is reproduced (see Tab. 1). The temporal evolution of the (optical) bolometric luminosity of SN 2002ap was roughly $\propto t^{0.8}$ before 10 days since the explosion, and $\propto t^{-0.5}$ after that until 20 days \citep{yoshii2003,tomita2006}. Thus, if the IC cooling dominates, then the radio temporal evolution will follow $L_{\nu} \propto t^{0.3}$ and $\propto t^{-1}$ before and after the optical peak (if the frequency of interest is optically thin). Well after the optical peak, the cooling will become less important as time goes by, so the radio light curve will be flattened until it reaches to the temporal behavior of $L_{\nu} \propto t^{-0.8}$ (for $p=2$). Thus, this scenario explains roughly the radio flux evolution, $\beta \sim -0.9$ as was observed. 

Although the scenario is consistent with the available observational data and also has a strength to explain the X-ray emission together with the radio \citep{bjornsson2004}, a drawback of the scenario is that it requires the electron distribution quite different from other SNe Ib/c. Note that for the magnetic field strength of $B \sim 0.3$ gauss \citep{bjornsson2004}, the radio emission is produced by electrons with the energy $\gamma \sim 50 - 150$, similar to those responsible for radio emission from other SNe cIIb/Ib/Ic \citep[e.g.,][]{maeda2012}. Thus, the different slope cannot be attributed to different energy regimes probed for different objects. If this is true, it might mean that being a broad-line SN Ic, SN 2002ap might have much more efficient electron acceleration \citep{chevalier2006b}, but as we show in \S 2 the same argument should apply to SN 2007gr, which is a non-broad line canonical SN Ic, making this interpretation less appealing. 

In this paper, we suggest another solution for radio emission from SN Ic 2002ap, which also naturally explains why `canonical' SN Ic 2007gr shares the similar properties. The analysis in the previous section suggests that the ejecta expansion of SNe 2002ap and 2007gr is approximated by the free expansion without any deceleration, rather than by the decelerating self-similar solution. In this case, $\alpha \sim -1$ and $\beta \sim -1$ are obtained with $p = 3$, if the cooling is not efficient. This is the same situation as has been inferred for other SNe cIIb/Ib/Ic \citep[e.g.,][]{chevalier2006b,soderberg2012}, with the only difference in the dynamic evolution: Assuming the general value of $p = 3$, then we predict that SNe in the low-density environment peak early in radio with the temporal index of $\beta \sim -1$, while in the higher density SNe peak later and show the temporal evolution of $\beta \sim -1.3$. 

The radio light curves computed for the free expansion case as compared to those of SN 2002ap are shown in Figure 6. The microphysics parameters are set as $\epsilon_{B} = \epsilon_{\rm e} = 0.1$ as typically assumed for radio SNe. As expected from Figure 5, we require relatively high forward shock velocity ($V_{\rm sh} = 60,000$ km s$^{-1}$) and low density CSM ($A_{*} = 0.007$). The synchrotron cooling is included in the model, but its effect is negligible. Thus, the light curves show the temporal slope $\beta \sim -1$ as observed (red-solid line in the figure), if the other cooling mechanism, i.e., the IC cooling, is ignored. 

Although it looks like a simple and straightforward interpretation, a story is complicated with the effect of the IC cooling. In Figure 6, we also show the light curves with the IC cooling included. With the set of parameters adopted, the IC cooling is indeed important, confirming the claim by the previous works \citep{bjornsson2004,chevalier2006b}. With the standard $p \sim 3$, the model now fails to reproduce the spectral index. 

Since the match between the observed radio behaviors and the prediction without the IC cooling is striking, we investigate what conditions will be necessary to remedy the problem. The Compton cooling time scale is larger if $\epsilon_{\rm e}$ is smaller to reproduce a given luminosity. Figure 7 shows an example where we adopt a low value for $\epsilon_{\rm e}$. The model parameters are $V_{\rm sh} = 80,000$ km s$^{-1}$, $A_{*} = 0.05$, $\epsilon_{B} = 0.1$, and $\epsilon_{\rm e} = 10^{-3}$. The required mass loss parameter is now increased roughly following the SSA scaling, $\epsilon_{B} A_{*} (\epsilon_{\rm e}/\epsilon_{B})^{8/19}$, but it is still very low as compared to other SNe Ib/c. The required velocity is also increased for the smaller value of $\epsilon_{\rm e}$ roughly following the SSA scaling $V_{\rm sh} \propto (\epsilon_{\rm e}/\epsilon_{B})^{-1/19}$. Not only the decrease in the relativistic electron density, but also the increase in the velocity, thus radius, has an effect to reduce the IC cooling effect. The light curve with low $\epsilon_{\rm e}$ is similar to the standard case. The difference is that in this model, the IC cooling is now negligible.  With the parameters for this `inefficient electron acceleration' model and $E_{51} = 5$ and $M_{\rm ej} = 3 M_{\odot}$, equation 5 predicts that $t_{\rm dec} \sim 360$ days, justifying our assumption of the free expansion. 

For SN 2007gr, we have found the similar solution with SN 2002ap (Figures 8 \& 9). The radio emission for the free expansion evolution explains the multi-band light curves fairly well without the IC cooling. With $\epsilon_{\rm e} = \epsilon_{B} = 0.1$ again the IC cooling alters the spectral index, and we require a low value of $\epsilon_{\rm e}$ to fit the radio properties. For our fiducial model ($V_{\rm sh} = 70,000$ km s$^{-1}$, $A_{*} = 0.15$, $E_{51} = 2$, and $M_{\rm ej} = 2 M_{\odot}$), $t_{\rm dec} \sim 40$ days, which is long and roughly consistent with the observation. 

We note that the low value of $\epsilon_{\rm e}$ is not physically unreasonable. Indeed, detailed modeling of radio properties of SN eIIb 1993J \citep{fransson1998} and SN cIIb 2011dh \citep{maeda2012} point to $\epsilon_{\rm e} \lsim 0.01$, lower than generally assumed. No strong constraints have been placed for other SNe \citep[e.g.,][]{chevalier2006a}. We discuss a few issues from the previous works on $\epsilon_{\rm e}$ in Appendix. The low efficiency of the electron acceleration has an important consequence in interpreting the X-ray emission, which is also discussed in Appendix. There, we conclude that so far there is no strong observational indication against the low value of $\epsilon_{\rm e}$ in stripped-envelope SNe (and we suggest that the low value of $\epsilon_{\rm e}$ may be a generic feature).

\section{Implications for shock breakout and progenitors}

\subsection{SNe Ic 2002ap and 2007gr: Structure of WR Stars at The End of Their Lives}

Following our interpretation that the radio emission from SNe showing a short time scale to a low radio peak luminosity is described by the free-expansion of the SN ejecta without deceleration, we can potentially connect properties of the shock breakout and the progenitor to these early time radio properties. In this early epoch, if the forward shock is still in the free-expansion phase, the characteristic velocity obtained through the radio properties is related to the maximum velocity of the SN ejecta following the shock breakout -- thus, this important information can be obtained through the radio properties in the first month after the explosion, although in the other wavelengths it requires to catch the very moment of the shock breakout. We emphasize that this is not possible for SNe already in the self-similar decelerating phase. Once the dynamics enters into the self-similar phase, the forward shock velocity is no more related to the initial maximum velocity of the ejecta (thus the shock breakout velocity) (see equation 4, also Fig. 2).\footnote[3]{It is much the same as the Sedov solution, where the information on the explosion is lost except for the explosion energy and the environment density.} If there would be exactly the same two SNe except for the maximum velocity ($V_{\rm SN}$), discriminating these SNe requires information at $t \lsim t_{\rm dec}$. 

According to \S 2 (Figures 2, 3, \& 4), the freely expanding forward shock velocity is related to the maximum velocity of the SN ejecta after the shock breakout. Adopting the parameters derived by the free-expansion model for the SN 2002ap radio emission, we have performed the same hydrodynamic simulation as in \S 2. We thereby find that $V_{\rm sh} = 80,000$ km s$^{-1}$ is obtained if $V_{\rm SN} \sim 90,000$ km s$^{-1}$. In this simulation, the forward shock speed is constant until $t_{\rm d} \sim 40$, thus being consistent with the free-expansion model for the radio emission. We have done the same experiment for SN 2007gr. Adopting $A_{*} = 0.15$, indeed we find it is not possible to obtain the constant forward shock velocity of $V_{\rm sh} \sim 70,000$ km s$^{-1}$ for more than 10 days, thus this model (with $\epsilon_{\rm e} = 10^{-3}$) for the radio emission is inconsistent with the dynamic evolution. However, if we adopt $A_{*} = 0.05$ (that corresponds to the required parameters of $\epsilon_{\rm e} \sim 0.01$ and $V_{\rm sh} \sim 60,000$ km s$^{-1}$ for $\epsilon_{\rm B} = 0.1$), then the constant forward shock velocity of $V_{\rm sh} \sim 60,000$ km s$^{-1}$ is obtained for $t_{\rm d} \sim 50$ for the maximum ejecta velocity of $V_{\rm SN} \sim 75,000$ km s$^{-1}$, being roughly consistent with the radio observation. 

Thus we infer the maximum ejecta velocity (or the `post-shock breakout velocity') of $\sim 0.25 - 0.3c$ for SNe Ic 2002ap and 2007gr through their early radio emissions. This is greater than inferred for SN 1987A, indicating that the progenitors of these SNe Ic are more compact than that of SN 1987A and/or the ejecta masses are smaller. However, the inferred velocity is smaller than the shock breakout velocity expected from the WR progenitors for SNe Ib/c (especially the WO or WC progenitor for SNe Ic). If $R_{*} \sim 5 R_{\odot}$, one expects $V_{\rm SN} \sim 0.35 c$ for typical SNe Ib/c, and $\sim 0.6 c$ for the ejecta properties of SN 2007gr, $\sim 0.8 - 0.9 c$ for SN 2002ap (adopting $\kappa \sim 0.2$ cm$^{2}$ g$^{-1}$ for He or C+O composition). This could mean that there is still something missing in our understanding of the SN Ic progenitors and/or there is a missing part in our understanding of the shock breakout dynamics. In other words, the apparent discrepancy would provide a hint on these still-unresolved issues, highlighting the importance of the independent information we could obtain through the early radio emission. We suggest the apparent discrepancy does provide an information on the progenitor structure just before the explosion. 

A possible solution we suggest is the following. While properties of typical Galactic WR stars are assumed in studying the SN explosion properties (including the shock breakout), we are not yet sure about what are properties of the WR star in a short period of time near the end of their lives. It has been suggested that the envelope would become expanded to a few to 10 times the original core radius, as a star approaches to the Eddington luminosity \citep{grafener2012}. Studying SN properties so far has not provided any confirmation on this picture, and the new insight we obtain here could be the first indication that such an evolution could be the case. 

\citet{grafener2012} showed that the typical density in such an envelope is $\sim 10^{-10}$ g cm$^{-3}$, and $\sim 10^{-8} M_{\odot}$ is contained within $\sim 10 R_{\odot}$. The envelope can be more massive depending on the WR mass and other parameters, reaching to $\gsim 10^{-6} M_{\odot}$ in the model sequence studied by \citet{grafener2012}. It is likely that this envelope would not affect the shock breakout itself -- adopting $E_{51} = 1$, $M_{\rm ej} = 3 M_{\odot}$, and $R_{*} = 5 R_{\odot}$, the stellar density at the shock breakout is estimated to be $\sim 10^{-8}$ g cm$^{-3}$ using the formalisms from \citet{rabinak2011}.\footnote[4]{However, the envelope could dilute and delay the shock breakout signal, as is similar to the situation expected for a shock breakout within a dense wind \citep{soderberg2008,chevalier2008}.} Thus, this would not dramatically affect basic predictions for the shock breakout high energy signal and subsequent early optical/UV emission, consistent with a few constraints favoring compact progenitors through the shock breakout X-ray \cite[SN Ib 2008D: ][]{soderberg2008,modjaz2009} and through the post-shock breakout optical emission \citep[SN Ic PTF10vgv: ][]{corsi2012}. 

However, the envelope could affect the dynamics just after the shock breakout. For typical parameters for SNe Ic, the shock breakout is estimated to take place when the mass above the shock (in the outer power law part, thus excluding the envelope here) is $\sim 10^{-8} M_{\odot}$. This is indeed comparable to the envelope mass, or can be even smaller. Then, the highest-velocity ejecta will experience deceleration during penetrating into the envelope, before the velocity profile is frozen when the shock emerges out of the envelope. We suggest this is what we infer from the early radio emission. If the envelope mass is $\sim 10^{-6}M_{\odot}$, it would decelerate the highest velocity to a half of the original shock breakout velocity, following $V_{\rm SN}/V_{\rm SN, 0} \propto (10^{-6} M_{\odot}/10^{-8} M_{\odot})^{-0.15}$ (where $V_{\rm SN, 0}$ is the shock breakout velocity, while $V_{\rm SN}$ is the maximum velocity after being affected by the envelope). This relation was obtained by combining the equations presented by \citet{rabinak2011}. The envelope is probably required to be He-rich to make the envelope inflation, but this amount of He is much smaller than the upper limit obtained by spectral modeling of SNe Ic \citep{hachinger2011}.

\subsection{Implications for Other classes of SNe}

The `engine-driven' SNe linked to GRBs \citep{kulkarni1998,soderberg2006,soderberg2010b} are distinguished from other stripped-envelope SNe in radio properties \citep{berger2003,soderberg2006b}. According to Fig. 5, these radio-strong SNe are roughly consistent with arising from the forward shock wave following the self-similar expansion, with the SN highly energetic ($E_{51} \sim 50$). The observed luminosity is larger than the expectation nearly an order of magnitude, but we note that in this regime the relativistic treatment is necessary for quantitative comparison. The rough agreement suggests that the strong radio emission from these SNe may be understood in terms of the standard SN-CSM interaction scenario (with the relativistic ejecta which would require the `central engine'), rather than invoking an additional relativistic jet component as long as the radio emission is concerned. The distinguished feature in these engine-driven SNe is the large explosion energy, which has been derived through the optical modeling \citep[e.g.,][]{iwamoto1998,woosley1999,maeda2006}. 

A majority of SNe Ib/c, and cIIb are consistent with the self-similar expansion in the $t_{\rm p} - L_{\rm p}$ plot. First, this indicates that their ejecta masses are mostly in the range of $M_{\rm ej} = 1 - 3 M_{\odot}$. An important implication is that then a majority of them are likely an explosion of stars with $M_{\rm ms} < 25 M_{\odot}$ and require the binary companion to strip off their H-rich envelope: For example, if $M_{\rm ms} \sim 25 M_{\odot}$ (roughly a lower limit for a single massive progenitor for SNe Ib/c), then the ejecta mass is expected to be $\sim 4.5 M_{\odot}$ if it explodes as an SN Ic and $\sim 6.5 M_{\odot}$ if an SN Ib. Next, this sets a rough upper limit for the size of the progenitors for these SNe cIIb/Ib/Ic. To enter into the self-similar phase, their shock breakout velocity must exceed at least $0.1 c$. For a reference value of $E_{51} = 1$ and $M_{\rm ej} = 3 M_{\odot}$, this means that the progenitor radius of most, if not all, of SNe cIIb/Ib/Ic is $R_{*} \lsim 250 R_{\odot}$.\footnote[5]{The upper limit here would further go down if the structure of the progenitors of these SNe is also similar to what we suggest for SNe 2002ap and 2007gr.} Thus, we reject an RSG progenitor for SNe cIIb/Ib/Ic -- they must come from a WR star, or at most a yellow giant \citep[the latter is a possible progenitor for SN cIIb 2011dh: ][]{maund2011,vandyk2011,bersten2012,benvenuto2012}. 

Another expectation is that some of SNe eIIb (`extended' SNe IIb) may follow the free-expansion for a long time during the phases when the radio observations are performed, because of the expected low shock breakout velocity. With $R_{*} \sim 500 R_{\odot}$, $E_{51} \sim 1$, and $M_{\rm ej} \sim 3 M_{\odot}$, SN eIIb 1993J is expected to have had the shock breakout velocity of $26,000$ km s$^{-1}$. Taking this as a face value for SNe eIIb, we expect that $t_{\rm dec} \sim 600, 60, 6$ day for $A_{*} = 1, 10, 100$. If there are SNe eIIb within a relatively low CSM environment (e.g., $A_{*} = 10$ for $\dot M = 10^{-6} M_{\odot}$ yr$^{-1}$ and $v_{\rm w} = 10$ km s$^{-1}$), then such SNe eIIb should show the free expansion phase in radio, and the high frequency observations will be especially useful to catch the feature. Indeed, SNe eIIb tend to be below the `self-similar expectation' in the $t_{\rm p} - L_{\rm p}$ plot, inferring that some of them might be indeed in the free expansion phase. However, a complication is that in these cases the free-free absorption may become important, and either a more detailed model including the free-free absorption or the direct measurement of the shock velocity by the VLBI will be necessary.

\section{Conclusions and Discussion}

In this paper, we have investigated a consequence of SNe Ib/c exploding within low density CSM (i.e., $A_{*} \lsim 0.1$, or $\dot M \lsim 10^{-6} M_{\odot}$ yr$^{-1}$ for a typical WR wind velocity). Such an SN should show a `free expansion' phase before entering into the self-similar deceleration phase, if the post-shock breakout velocity is $V_{\rm SN} \lsim 0.3c$. The predicted radio properties are different from those SNe in the self-similar phase, characterized by a shallow decline in the temporal evolution while the spectral index is the same, if all the other characters (i.e., the acceleration mechanism of electrons) are unchanged. 

SNe exploding in the low density CSM environment should be characterized in radio frequency by the fast rise to the peak (within 10 days after the explosion) and the low luminosity there. We have shown that all stripped-envelope SNe with these properties observed so far (SNe 2002ap and 2007gr, as well as SN 1987A from a blue giant progenitor) indeed have the expected properties in the temporal and spectral indexes. The synthesized multi-band light curves show a good match to the observed ones for SNe 2002ap and 2007gr. We note that the other example, SN 1987A, was also well modeled by the free expansion dynamics \citep{chevalier1998}. In our interpretation, the efficiency of the acceleration of electrons must be low in order to avoid the IC cooling in radio frequencies. This is the same conclusion we obtained for SN 2011dh \citep{maeda2012}, and we suggest this may be a generic property shared by the SN-CSM interaction. 

Understanding the radio properties from SNe in the $t_{\rm p} - L_{\rm p}$ plot in terms of the dynamics of the shock propagation, we propose new diagnostics on the shock breakout and progenitor properties through the early radio emission: The forward shock velocity in the free expansion phase ($V_{\rm sh}$) is closely related to the maximum velocity obtained at/after the shock breakout ($V_{\rm SN}$). We suggest that the relatively low post-shock breakout velocity ($\sim 0.3c$) we have derived for SNe Ic 2002ap and 2007gr could indicate the existence of an  envelope driven by the near-Eddington luminosity near the end of the WR evolution \citep{grafener2012}. This highlights the usefulness/uniqueness of the proposed strategy as compared to other methods (e.g., breakout flash and/or subsequent optical emission) to probe the shock breakout and progenitor: the velocity information is not obtained by the other methods. Also, for a majority of SNe cIIb/Ib/Ic, we reject a red-supergiant progenitor through the radio properties. 

The idea can in principal provide slightly different approaches in estimating the post-shock breakout maximum velocity and placing constraints on progenitor structures. Once one identifies the free expansion phase (i.e., the transition from the free-expansion to the self-similar dynamics in the decay slope), then one could estimate the post-shock breakout velocity using equation (5) or similar expression, adopting ejecta properties and the CSM density estimated independently. The expression however requires further calibrations, and does not provide a cross check of the decay slope. For these reasons, we adopted a more detailed approach, in which we fitted the multi-band light curves and checked the assumed dynamic evolution with hydrodynamics simulations. Also, a stronger constraint on the progenitor radius when SNe do not show the free expansion phase than we did for most radio-detected SNe IIb/Ib/Ic (using the information at the radio peak) could be placed, by using the earliest data points in which SNe are in the self-similar phase. A complication is that either the absorption (in the low frequency) or cooling (in the high frequency) can change the temporal slope in the early phase, thus distinguishing the different dynamics is generally difficult well before the radio peak. For this reason, we have placed a conservative upper limit for the progenitor radius using the radio information around the peak. Further study and calibrations of these methods could be possible applications of the idea presented in this paper. 

Future, large observational data set in radio frequency will be highly valuable. Such data are increasingly accumulated recently thanks to great efforts of researchers working in the field \citep[e.g.,][]{soderberg2012}. SNe in the low density CSM environment like SNe 2002ap and 2007gr will provide a new possibility to tackle to properties of the shock breakout and progenitor as mentioned above. A majority of radio-detected SNe IIb/Ib/Ic follow the self-similar evolution, and thus their properties and distribution in the $t_{\rm p} - L_{\rm p}$ plot will tell general distribution of the progenitor mass and the energetics independently from optical wavelengths: So far, their distribution suggests that a majority of them (i.e., radio-detected stripped-envelope SNe) are from relatively low mass progenitor ($\lsim 25 M_{\odot}$), indicating that a binary interaction path could be a dominant path, in line with other recent studies \citep[e.g.,][]{sana2012}. 

The low density CSM around SNe Ic 2002ap and 2007gr may be related to the WR progenitor structure just before the explosion. If the progenitor luminosity is larger, it would likely produce a higher velocity wind, resulting in a lower value of $A_{*}$. This probably favors a massive WO star as a progenitor of these SNe Ic \citep[e.g.,][]{nugis2000}. If this is true, we expect no SNe IIb/Ib would belong to the `rapid and faint' radio stripped-envelope SNe (i.e., SNe 2002ap and 2007gr are both of type Ic), and future increasing samples should be able to address to this question.\footnote[6]{SN II 1987A may well be an exception, due to the low metallicity environment there. This suggests that detailed analysis will require to take into account the metallicity of the local environment as well.} Searching and observing these radio faint stripped-envelope SNe will provide new clues on the progenitor scenario from this aspect as well. Especially, this will be best done by future observatories with the better sensitivity than currently. 
Because of the low radio luminosity of these objects, there may well be an observational bias in which we underestimate the frequency of these radio-weak SNe \citep{soderberg2010}. Once a volume-limited sample is constructed, it will hopefully connect the radio properties with different progenitor evolutionally paths. 

\acknowledgements 
K.M. thanks the anonymous referee for many valuable comments and suggestions. K.M. also thanks Claes Fransson for his comments on the manuscript and stimulating discussion. This research is supported by World Premier International Research Center Initiative (WPI Initiative), MEXT, Japan and by Grant-in-Aid for Scientific Research for Young Scientists (23740141). K.M. acknowledges a support by the Stockholm University as a short-term visitor, and thanks Jesper Sollerman and the staffs there for their hospitality.


\clearpage
\begin{figure*}
\begin{center}
        \begin{minipage}[]{0.3\textwidth}
                \epsscale{1.0}
                \plotone{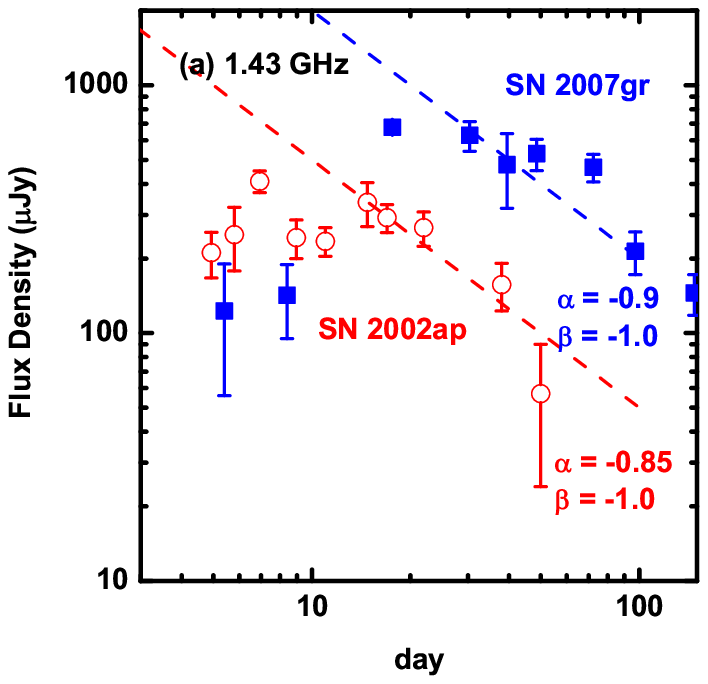}
        \end{minipage}
        \begin{minipage}[]{0.3\textwidth}
                \epsscale{1.0}
                \plotone{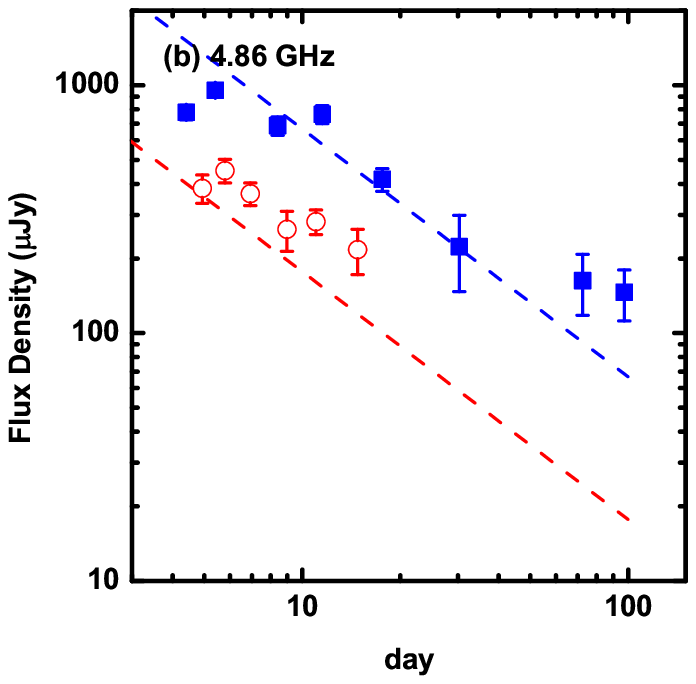}
        \end{minipage}
        \begin{minipage}[]{0.3\textwidth}
                \epsscale{1.0}
                \plotone{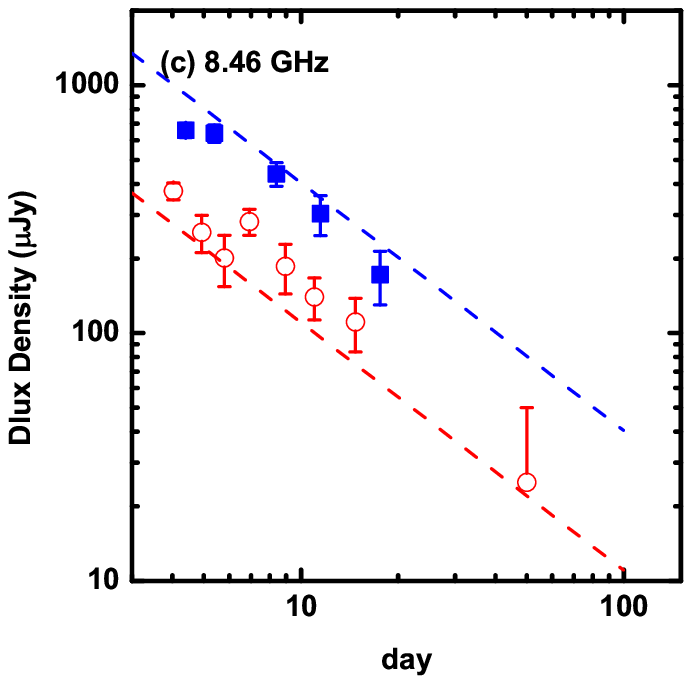}
        \end{minipage}
\end{center}
\caption
{Observed radio light curves of SNe 2002ap and 2007gr \citep{berger2002,soderberg2010}. The behavior, $L_{\nu} \propto \nu^{\alpha} t^{\beta}$ as normalized at 1.43 GHz, 
is shown by lines ($\alpha = -0.85$ and $-0.9$ for SNe 2002ap and 2007gr, respectively, and 
$\beta = -1.0$ for both SNe). SNe 2002ap and 2007gr show similar light curves, both in the spectral index and temporal slope, in the optically thin radio emission. Distances of 8 Mpc and 9.3 Mpc are adopted for SNe 2002ap and 2007gr, respectively. 
\label{fig_radio_obs}}
\end{figure*}

\begin{figure*}
\begin{center}
        \begin{minipage}[]{0.6\textwidth}
                \epsscale{1.0}
                \plotone{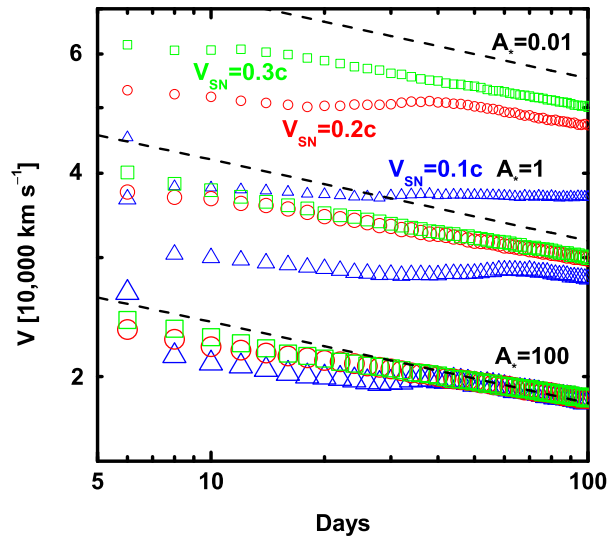}
        \end{minipage}
\end{center}
\caption
{Evolution of the forward shock velocity computed by numerical hydrodynamic simulations, for $V_{\rm SN} = 0.3 c$ (green open squares), $0.2 c$ (red open circles), and $0.1 c$ (blue open triangles). The CSM density parameter is $A_{*} = 0.01$ (small symbols, top), $A_{*} = 1$ (medium-sized symbols, middle), and $A_{*} = 100$ (large symbols, bottom). The self-similar solution is shown by dotted lines for different values of $A_{*}$. 
\label{fig_vel_t}}
\end{figure*}

\clearpage
\begin{figure*}
\begin{center}
        \begin{minipage}[]{0.6\textwidth}
                \epsscale{1.0}
                \plotone{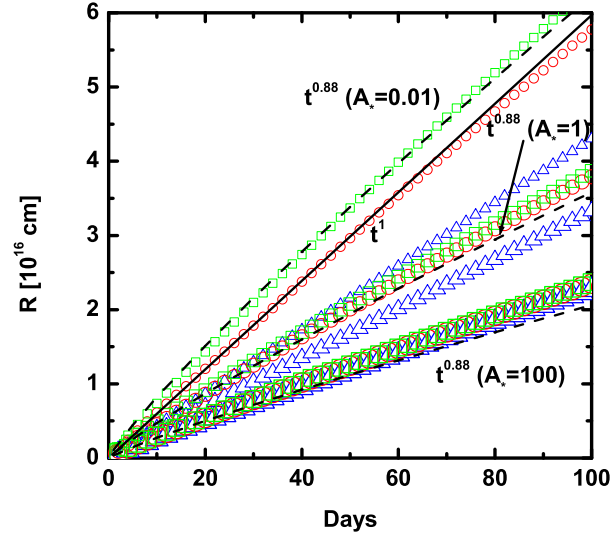}
        \end{minipage}
\end{center}
\caption
{Same as Figure 2, but for the evolution of the radius of the shock wave. See the caption of Figure 2 for meaning of different symbols. The self-similar solution is shown for different values of $A_{*}$ (dotted lines). The free-expansion evolution (i.e., $R \propto t$) is shown by a solid line that fits well the evolution of the shock with $V_{\rm SN} = 0.2 c$ and $A_{*} = 0.01$ until $t_{\rm d} \sim 70$. 
\label{fig_r_t}}
\end{figure*}

\begin{figure*}
\begin{center}
        \begin{minipage}[]{0.6\textwidth}
                \epsscale{1.0}
                \plotone{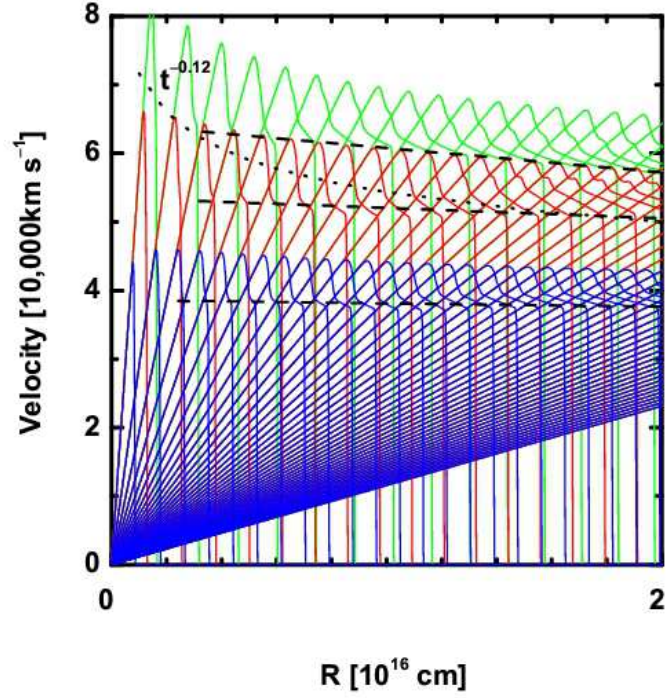}
        \end{minipage}
\end{center}
\caption
{Evolution of the velocity profiles at the beginning of the interaction, for the low density CSM with $A_{*} = 0.01$. Three models are shown, with $V_{\rm SN} = 0.3 c$ (green), $0.2 c$ (red), and $0.1 c$ (blue). The position of the forward shock is indicated by the dashed lines, while the behavior expected from the decelerated self-similar expansion is shown by the dotted lines (with the vertical scale arbitrary). The model profiles are shown starting at $t_{\rm d} = 1$, with the interval of $2$ days in different snap shots. 
\label{fig_hyd_r_v}}
\end{figure*}

\clearpage
\begin{figure*}
\begin{center}
        \begin{minipage}[]{0.95\textwidth}
                \epsscale{1.0}
                \plotone{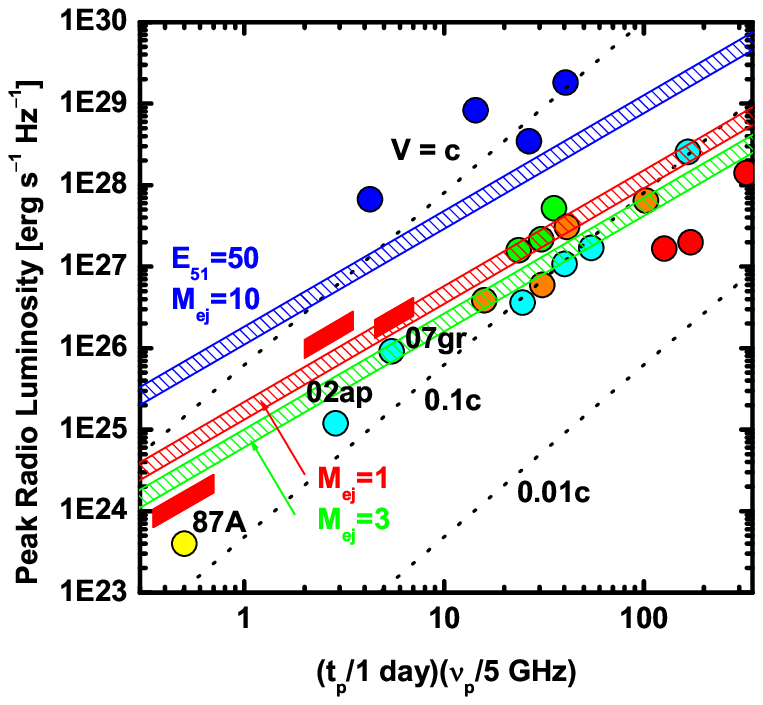}
        \end{minipage}
\end{center}
\caption
{Radio properties of SNe IIb/Ib/Ic and SN II 1987A in the $t_{\rm p} - L_{\rm p}$ plot. The observed points are reproduction of \citet{soderberg2012} \citep[see also][]{chevalier2006a,chevalier2010}, showing SNe Ic (cyan), Ib (green), cIIb (orange), eIIb (red), `engine-driven' SNe Ic (including those associated with GRBs: blue), and 1987A (yellow). The velocity estimate with the SSA scaling is shown by dotted lines. The expectation from the self-similar hydrodynamic evolution is shown by the shaded area (blue, red, and green) for $\epsilon_{B} = 0.1$ and $\alpha \equiv \epsilon_{\rm e}/\epsilon_{B} = 0.1 - 1$, with $E_{51}$ and $M_{\rm ej}$ given in labels. The self-similar expectation for specific cases of SNe 2002ap ($E_{51} = 5$, $M_{\rm ej}= 3 M_{\odot}$), 2007gr ($E_{51} = 2$, $M_{\rm ej} = 2 M_{\odot}$), and 1987A ($E_{51} = 1.4$, $M_{\rm ej} = 14$) are shown by red-filled area. 
\label{fig_tplp}}
\end{figure*}

\clearpage
\begin{figure*}
\begin{center}
        \begin{minipage}[]{0.3\textwidth}
                \epsscale{1.0}
                \plotone{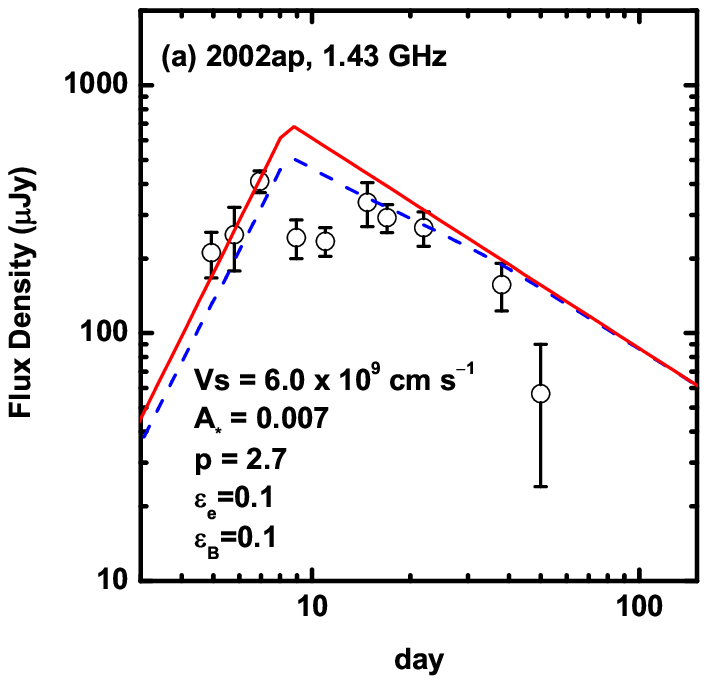}
        \end{minipage}
        \begin{minipage}[]{0.3\textwidth}
                \epsscale{1.0}
                \plotone{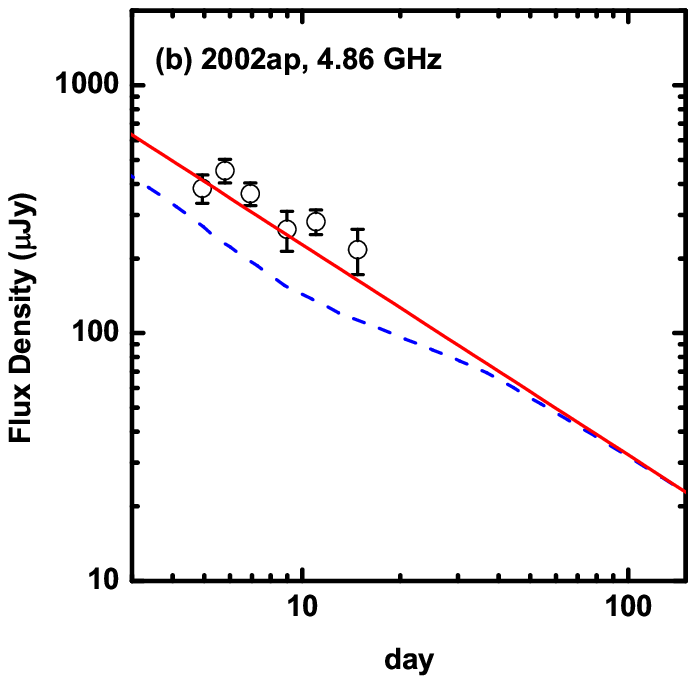}
        \end{minipage}
        \begin{minipage}[]{0.3\textwidth}
                \epsscale{1.0}
                \plotone{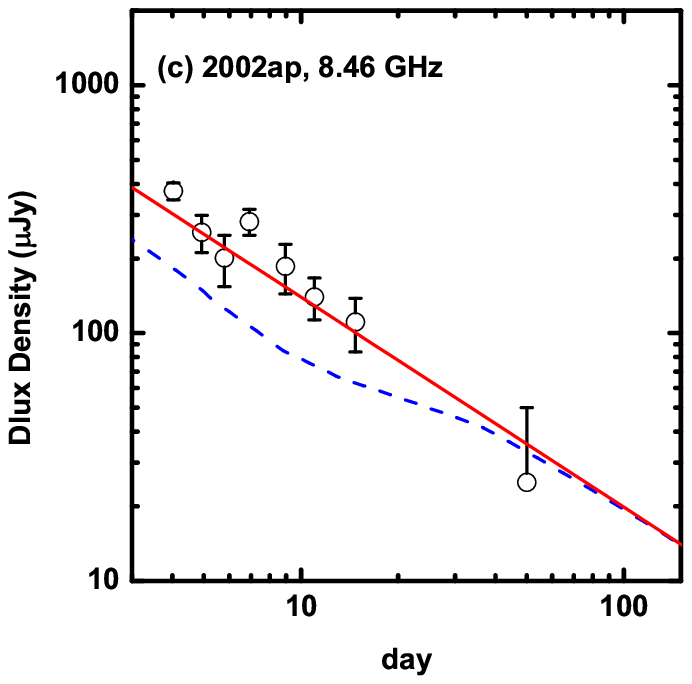}
        \end{minipage}
\end{center}
\caption
{Radio emission from SN 2002ap compared with the `efficient electron acceleration' model.  
The model adopts $V_{\rm sh} = 6.0 \times 10^9$ cm s$^{-1}$, $A_{*} = 0.007$, $p=2.7$, $\epsilon_{e} = \epsilon_{B} = 0.1$. The red-solid line is computed without the inverse Compton cooling, while the blue-dashed line includes the inverse Compton cooling with the bolometric light curve of SN 2002ap \citep{mazzali2002,yoshii2003}. The synchrotron cooling is included in both models, but its effect is negligible. 
\label{fig_2002ap_highe}}
\end{figure*}

\begin{figure*}
\begin{center}
        \begin{minipage}[]{0.3\textwidth}
                \epsscale{1.0}
                \plotone{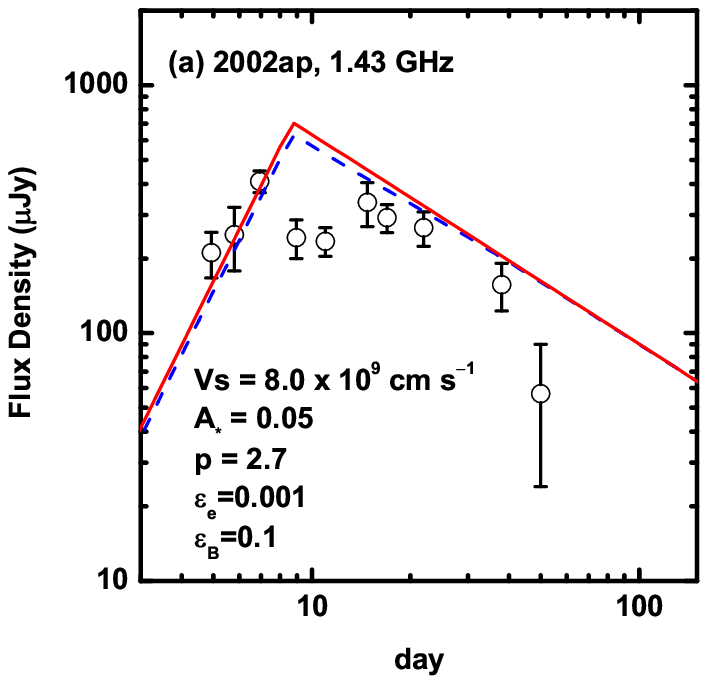}
        \end{minipage}
        \begin{minipage}[]{0.3\textwidth}
                \epsscale{1.0}
                \plotone{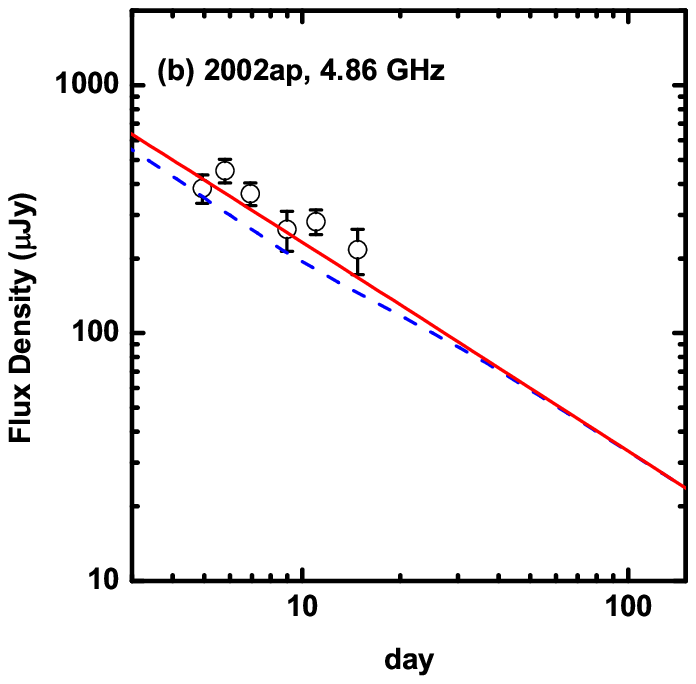}
        \end{minipage}
        \begin{minipage}[]{0.3\textwidth}
                \epsscale{1.0}
                \plotone{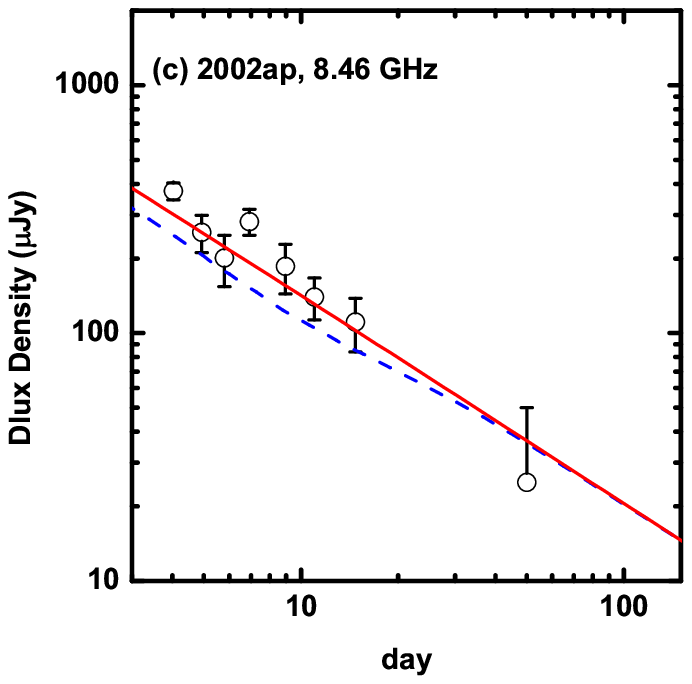}
        \end{minipage}
\end{center}
\caption
{Radio emission from SN 2002ap compared with the `inefficient electron acceleration' model.  
The model adopts $V_{\rm sh} = 8.0 \times 10^9$ cm s$^{-1}$, $A_{*} = 0.05$, $p=2.7$, $\epsilon_{e} = 10^{-3}$, $\epsilon_{B} = 0.1$. See the caption of Figure 6. 
\label{fig_2002ap_lowe}}
\end{figure*}

\clearpage
\begin{figure*}
\begin{center}
        \begin{minipage}[]{0.3\textwidth}
                \epsscale{1.0}
                \plotone{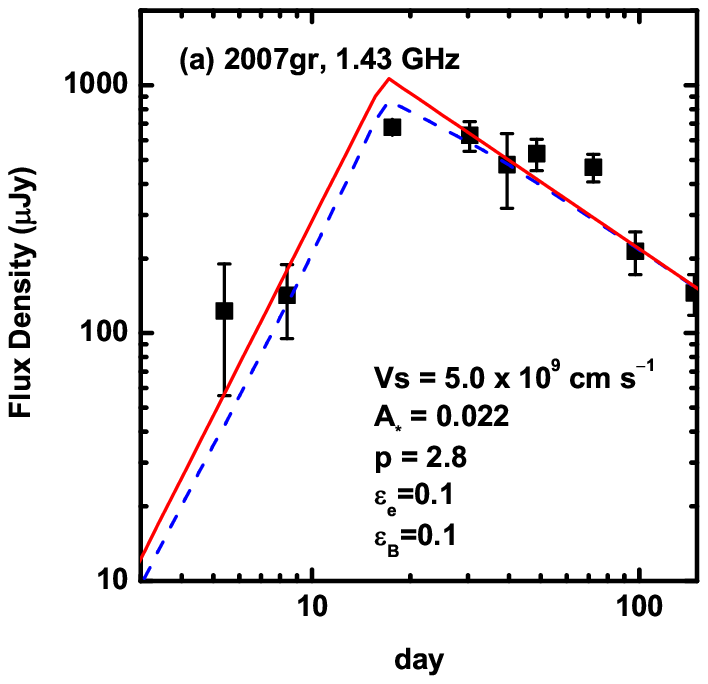}
        \end{minipage}
        \begin{minipage}[]{0.3\textwidth}
                \epsscale{1.0}
                \plotone{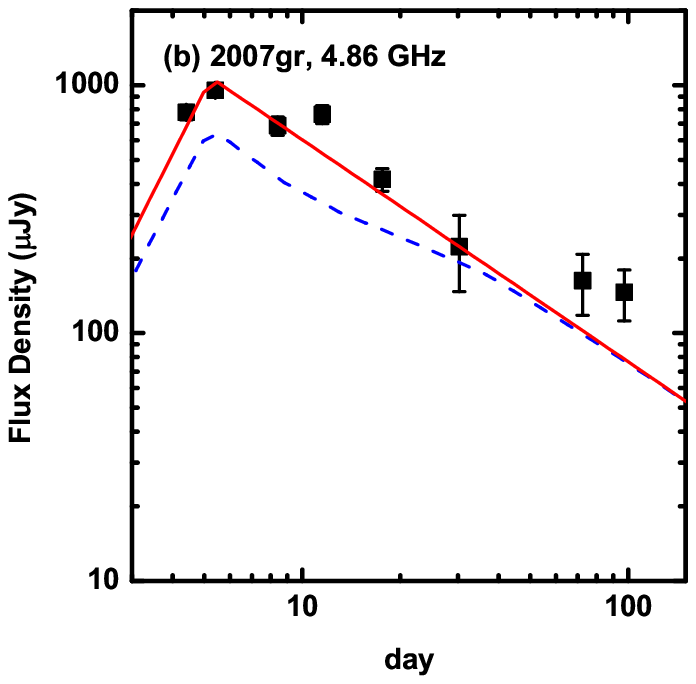}
        \end{minipage}
        \begin{minipage}[]{0.3\textwidth}
                \epsscale{1.0}
                \plotone{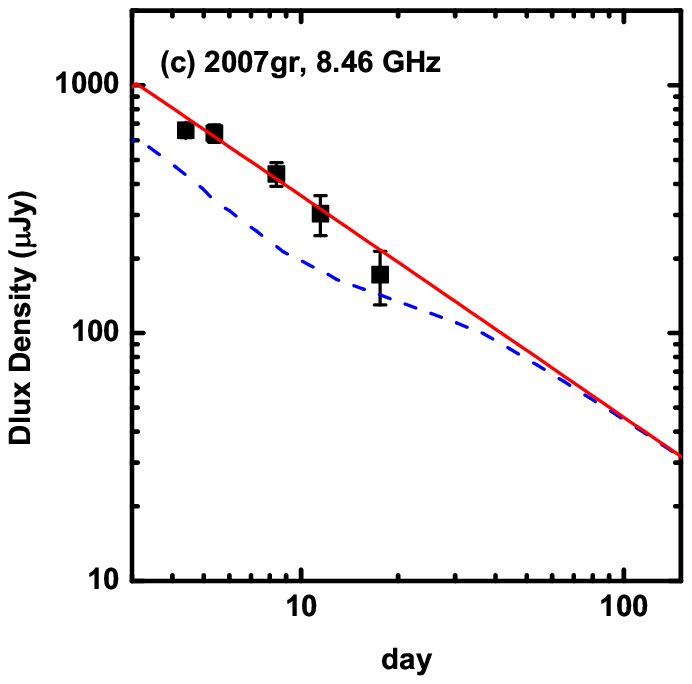}
        \end{minipage}
\end{center}
\caption
{Radio emission from SN 2007gr compared with the `efficient electron acceleration' model.  
The model adopts $V_{\rm sh} = 5.0 \times 10^9$ cm s$^{-1}$, $A_{*} = 0.022$, $p=2.8$, $\epsilon_{e} = \epsilon_{B} = 0.1$. See the caption of Figure 6. 
\label{fig_2007gr_highe}}
\end{figure*}

\begin{figure*}
\begin{center}
        \begin{minipage}[]{0.3\textwidth}
                \epsscale{1.0}
                \plotone{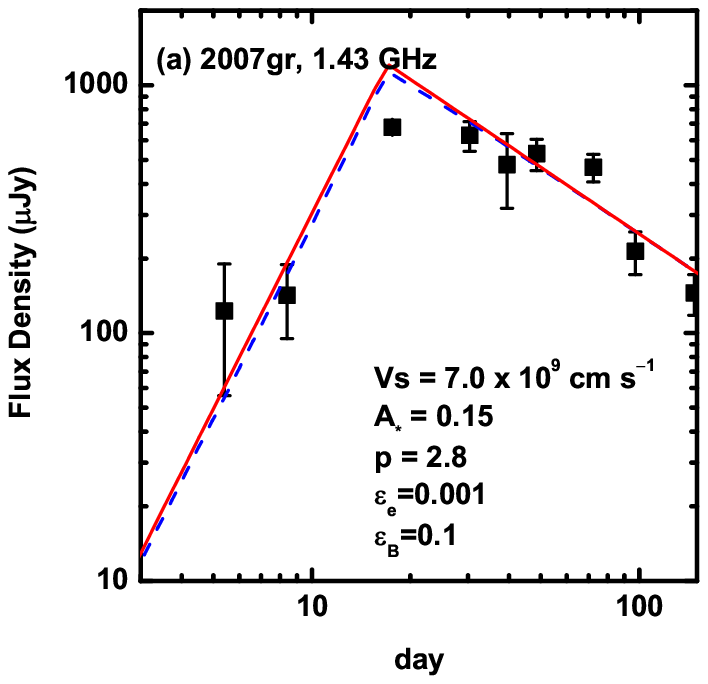}
        \end{minipage}
        \begin{minipage}[]{0.3\textwidth}
                \epsscale{1.0}
                \plotone{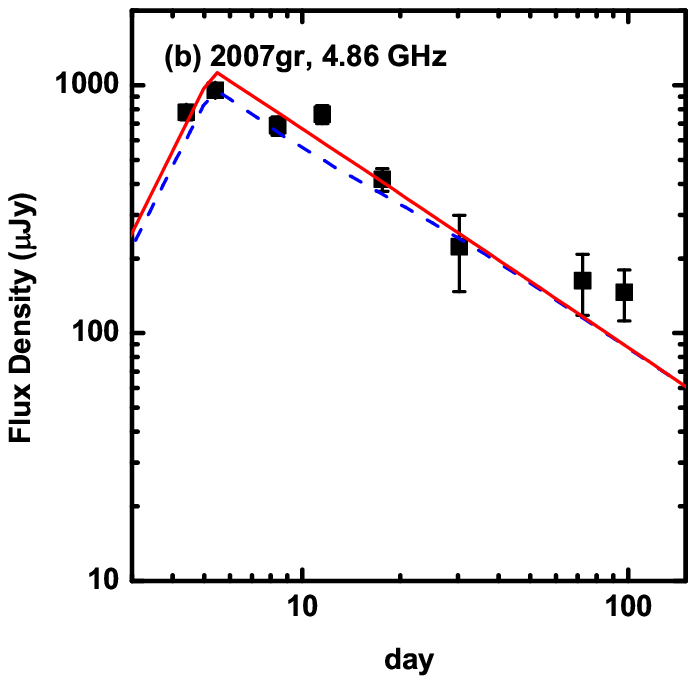}
        \end{minipage}
        \begin{minipage}[]{0.3\textwidth}
                \epsscale{1.0}
                \plotone{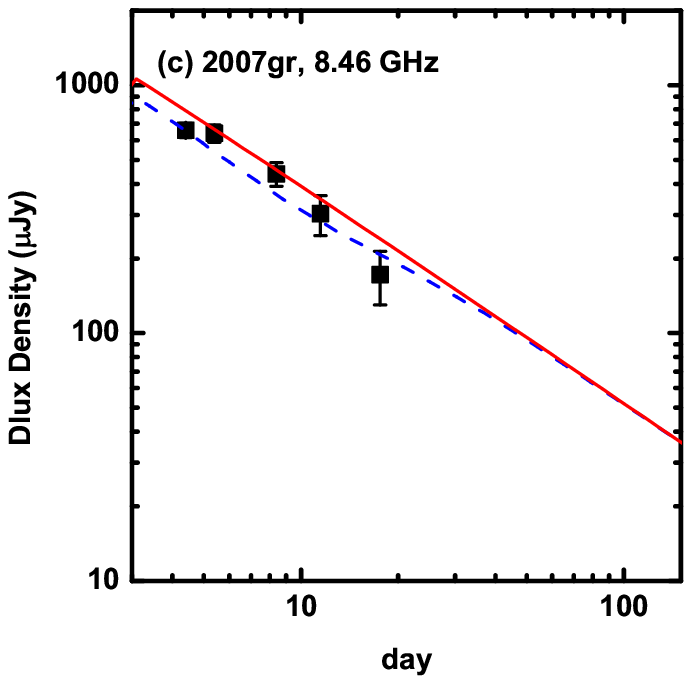}
        \end{minipage}
\end{center}
\caption
{Radio emission from SN 2007gr compared with the `inefficient electron acceleration' model.  
The model adopts $V_{\rm sh} = 7.0 \times 10^9$ cm s$^{-1}$, $A_{*} = 0.15$, $p=2.8$, $\epsilon_{e} = 10^{-3}$, $\epsilon_{B} = 0.1$. See the caption of Figure 6. 
\label{fig_2007gr_lowe}}
\end{figure*}

\clearpage
\begin{deluxetable}{cccc}
 \tabletypesize{\scriptsize}
 \tablecaption{Characteristics of the synchrotron emission\tablenotemark{a}
 \label{tab:slopes}}
 \tablewidth{0pt}
 \tablehead{
   \colhead{}
 & \colhead{Adiabatic}
 & \colhead{Syn.}
 & \colhead{IC}
}
\startdata
$\alpha$ & $\frac{1-p}{2}$            & $-\frac{p}{2}$  & $-\frac{p}{2}$ \\
$\beta$  & $(3m-3)+\frac{1-p}{2}$ & $(3m-3)+\frac{2-p}{2}$ & $(5m-5) +\frac{2-p}{2}+\delta$ \\\hline
$\alpha (p=2)$ & $-\frac{1}{2}$           & $-1$ & $-1$ \\
$\beta (p = 2)$ & $(3m-3)-\frac{1}{2}$ & $(3m-3)$  & $(5m-5)+\delta$ \\\hline
$\alpha (p=3)$ & $-1$         & $-\frac{3}{2}$ & $-\frac{3}{2}$ \\
$\beta (p=3)$ & $(3m-3)-1$ & $(3m-3)-\frac{1}{2}$ & $(5m-5)-\frac{1}{2}+\delta$\\
\enddata
\tablenotetext{a}{The spectral index ($\alpha$) and temporal slope ($\beta$) are shown for different cooling regimes ($L_{\nu} \propto \nu^{\alpha} t^{\beta}$). These are characterized by the electron distribution power law index ($p$), the evolution of the forward shock ($m$, where $R_{\rm sh} \propto t^{m}$), and the evolution of the bolometric luminosity ($\delta$, where $L_{\rm bol} \propto t^{\delta}$). }
\end{deluxetable}

\appendix

\section{Electron Acceleration Efficiency ($\epsilon_{\rm e}$)}

In our scenario, the acceleration of the relativistic electrons is required to be (relatively) inefficient ($\epsilon_{\rm e} \lsim 0.01$), as compared to frequently assumed ($\epsilon_{\rm e}\sim 0.1$). As mentioned in the main text, so far there is no strong observational constraint against the low value of $\epsilon_{\rm e}$. Indeed, \citet{fransson1998} and \citet{maeda2012} suggested $\epsilon_{\rm e} \lsim 0.01$ for SN eIIb 1993J and SN cIIb 2011dh, respectively, based on detailed modeling of their radio properties. We discuss this issue in this Appendix. 

\subsection{Previous Works}
It has been sometime stated that $\epsilon_{\rm e} \gsim 0.1$ for SNe Ib/c based on the equation $\epsilon_{\rm e} m_{\rm p} V_{\rm sh}^2 \sim \gamma_{\rm m} m_{\rm e} c^2$ where $\gamma_{\rm m} \gsim 1$ is the characteristic Lorentz factor of relativistic electrons \citep{soderberg2005,soderberg2010}. This is an analog of what is frequently adopted in the field of Gamma-Ray Burst non-thermal emission where the relativistic shock wave could accelerate all or most of the electrons into the relativistic speed. The assumption, that all the electrons are accelerated to the relativistic speed, however does not have to be the case in a non-relativistic shock wave in SNe. If only a fraction of $\zeta_{\rm e}$ (in number) are accelerated from the thermal population to relativistic energy, then one has to multiply the R.H.S of the above equation by $\zeta_{\rm e}$, and therefore this limit on $\epsilon_{\rm e}$ (by $\gamma_{\rm m} \gsim 1$) becomes lower by the same factor of $\zeta_{\rm e}$. This is taken into account in our formalism (equation 8), and this is explicitly expressed as $\zeta_{\rm e} \sim 4 \epsilon_{{\rm e}, -1} (V/0.1c)^2$ (from equations 3 and 8) for an H-rich CSM (of course it applies only for $\zeta_{\rm e} \lsim 1$). For reference values of $\epsilon_{\rm e} = 10^{-3}$ and $V \sim 0.2c$, we have $\zeta_{\rm e} \sim 0.1 - 0.2$, i.e., about 10\% of thermal electrons are accelerated to the relativistic speed. 

This value of $\zeta_{\rm e}$ seems reasonable from circumstantial evidences: (1) $\zeta_{\rm e}$ has been typically found to be less than 1\% in SN remnants where $V_{\rm sh} \sim 0.01c$ \citep[e.g.,][]{bamba2003}. Thus we do not expect that the condition $\zeta_{\rm e} \sim 1$ must be met in the SN Ib/c shock wave which is also non-relativistic. (2) The X-ray emission from SN IIb 1993J (similar to SNe Ib/c in the shock velocity) is believed to be emitted from the thermal electrons, and it requires that the thermal electrons are the bulk population at the shock front, while the relativistic electrons occupy only a small fraction in number \citep{fransson1996,fransson1998}. In any case, the value of $\zeta_{\rm e}$ is still an open question (not only for SNe but also other astrophysical acceleration sites). 

Furthermore, the existence of thermal electrons as a bulk population would not produce inconsistency in terms of observed radio properties. The synchrotron radio emission is produced by electrons with $\gamma \gsim 50$ for typical magnetic field strength seen in SNe Ib/c shock front, leaving no observational signature for electrons with $\gamma \lsim 50$ (including the thermal population) in the radio band \citep[see][for detailed discussion]{maeda2012}. Indeed, it is observationally forbidden to have $\gamma_{\rm m} >> 1$: In this case the radio emission should show a characteristic spectral break \citep{soderberg2005,chevalier2006b,soderberg2010},  although such a break is not seen in the observation. In this sense, the radio observation provides the upper limit for $\epsilon_{\rm e}$, not the lower limit.

\subsection{X-ray Production through the Inverse Compton Mechanism} 
The inverse Compton (IC) up-scattering of SN optical photons has been proposed as one of the X-ray emission production mechanism, especially favored for stripped envelope SNe with $A_{*} \sim 1$ \citep[e.g., ][]{bjornsson2004,chevalier2006b,soderberg2012}. The scenario requires a large population of relativistic electrons with $\gamma \lsim 50$, and thus a large value of $\epsilon_{\rm e}$ ($\gsim 0.1$) if these electrons' energy distribution follows a power law extrapolated from the radio-synchrotron emitting electrons (typically with $\gamma \sim 50 - 200$) \citep[see,][for detailed discussion]{maeda2012}. In this section, we stress that having a large value of $\epsilon_{\rm e}$ ($\gsim 0.1$) is not a single solution to produce the IC X-rays, but there is an alternative interpretation in which the energy distribution of the IC emitting electrons does not follow the extrapolation from the synchrotron emitting electrons as suggested by \citet{maeda2012}. 

The low efficiency of the electron acceleration thus has an important consequence in interpreting the X-ray emission. Among several models suggested for the X-ray behavior of SN 2002ap, the IC scattering scenario seems the most plausible \citep{bjornsson2004,chevalier2006b}. However, if we adopt $p \sim 3$, then we want to avoid the Compton cooling effect in the radio frequency (\S 4). We can estimate the IC X-ray  luminosity in the form similar to the radio synchrotron emission: 
\begin{equation}
\nu L_{\nu} \sim \pi R^{2} V n_{\rm e} \gamma^{2-p} m_{\rm e} c^2 \left(1+\frac{t_{\rm d}}{t_{\rm ic} (\gamma)}\right) \ , 
\end{equation}
where $t_{\rm ic}$ is the IC cooling time scale. The typical electron energy ($\gamma_{\rm ic}$) where the Compton effect is significant is not sensitive to the value of $\epsilon_{\rm e}$ since $\gamma_{\rm ic} \propto L_{\rm SN}^{-1} R_{\rm sh}$ (not dependent on the microphysics parameters): At $t_{\rm d} = 5$, $\gamma_{\rm ic} \sim 140$ and $250$ for our models with $\epsilon_{\rm e} = 0.1$ and $10^{-3}$, respectively. On the other hand, the electron energy responsible for the X-ray emission through the IC is $\gamma \sim 30$. Thus it is in the adiabatic phase, and the predicted X-ray luminosity is $\nu L_{\nu} (1 keV) \sim 6 \times 10^{36}$ erg s$^{-1}$ (for $\epsilon_{\rm e} = 0.1$) and $\sim 6 \times 10^{35}$ erg s$^{-1}$ (for $\epsilon_{\rm e} = 10^{-3}$) at day 5. The former is lower than the observed value only by a factor of two, but the latter is more than an order of magnitude lower than observed. The estimate is consistent with the previous work by \citet{bjornsson2004}. 

We note however that the similar situation is found in SN cIIb 2011dh, for which the intensive radio and X-ray data allowed detailed modeling of its properties \citep{soderberg2012,krauss2012,bietenholz2012,horesh2012}. For SN 2011dh, there is almost no doubt that the power law index of the relativistic electrons emitting in the radio frequencies is $p \sim 3$, constrained by the late-time, optically thin light curves in multi bands \citep{soderberg2012}. While \citet{soderberg2012} suggested a high value of $\epsilon_{\rm e}$ and \citet{horesh2012} suggested even more efficient electron acceleration with $\epsilon_{\rm e}/{\epsilon_{B}} \sim 1000$,  
\citet{maeda2012} pointed out that little change in the spectral index was observed once the radio emission became optically thin, and suggested that the IC cooling effect must be negligible in radio and that the acceleration of the relativistic electrons must be inefficient ($\epsilon_{\rm e} \lsim 0.01$). In this interpretation of \citet{maeda2012}, the predicted IC emission in the X-ray was about one or two orders of magnitudes smaller than observed -- this is the same situation we found for SNe 2002ap and 2007gr, but with the constraints on the spectral energy index of the electrons ($p$) and the effect of the IC cooling much stronger than for SNe 2002ap and 2007gr. To remedy this problem, \citet{maeda2012} suggested there is a distinct population of relativistic electrons below $\gamma \sim 50$, with the number density (per $\gamma$) more than an order of magnitude larger than the extrapolation from the power law distribution for the radio-emitting, higher energy electrons, in order to explain the X-rays from SN 2011dh by the IC mechanism while still being consistent with the radio behaviors. We expect that the acceleration mechanisms is essentially the same in SN 2002ap, thus suggest that the same argument for SN 2011dh applies to SN 2002ap as well. Then the X-ray properties of SN 2002ap could be explained in the same manner.


\begin{thebibliography}{}

\bibitem[Arnett et al.(1989)]{arnett1989}
Arnett, W.D., Barcall, J.N., Kirshner, R.P., \& Woosley, S.E. 1989, ARA\&A, 27, 629

\bibitem[Bamba et al.(2003)]{bamba2003}
Bamba, A., Yamazaki, R., Ueno, M., \& Koyama, K. 2003, ApJ, 589, 827 

\bibitem[Benvenuto et al.(2012)]{benvenuto2012}
Benvenuto, O., et al. 2012, ApJ, submitted (arXiv:1207.6807)

\bibitem[Berger et al.(2002)]{berger2002}
Berger, E., Kulkarni, S.R., Frail, D.A., \& Soderberg, A.M. 2003, ApJ, 599, 408

\bibitem[Berger et al.(2003)]{berger2003}
Berger, E., Kulkarni, S.R., \& Chevalier, R.A. 2002, ApJ, 577, L5

\bibitem[Bersten et al.(2012)]{bersten2012}
Bersten, C.M., et al. 2012, ApJ, 757, 31 

\bibitem[Bietenholz et al.(2012)]{bietenholz2012}
Bietenholz, M.F., et al. 2012, ApJ, 751, 125

\bibitem[Bj\"ornsson \& Fransson(2004)]{bjornsson2004}
Bj\"ornsson, C.-I., \& Fransson, C. 2004, ApJ, 605, 823 

\bibitem[Blinnikov et al.(2000)]{blinnikov2000}
Blinnikov, S., Lundqvist, P., Bartunov, O., Nomoto, K., \& Iwamoto, K. 2000, ApJ, 532, 1132

\bibitem[Chevalier(1982)]{chevalier1982}
Chevalier, R.A., 1982, ApJ, 258, 790 

\bibitem[Chevalier(1998)]{chevalier1998}
Chevalier, R.A. 1998, ApJ, 499, 810 

\bibitem[Chevalier et al.(2006a)]{chevalier2006a}
Chevalier, R.A., Fransson, C., \& Nymark, T.K. 2006a, ApJ, 641, 1029

\bibitem[Chevalier \& Fransson(2006b)]{chevalier2006b}
Chevalier, R.A., \& Fransson, C. 2006b, ApJ, 651, 381 

\bibitem[Chevalier \& Fransson(2008)]{chevalier2008}
Chevalier, R.A., \& Fransson, C. 2008, ApJ, 683, L135

\bibitem[Chevalier \& Soderberg(2010)]{chevalier2010}
Chevalier, R.A., \& Soderberg, A.M. 2010, ApJ, 711, L40 

\bibitem[Corsi et al.(2012)]{corsi2012}
Corsi, A., et al. 2012, ApJ, 747, 5

\bibitem[Falk \& Arnett(1977)]{falk1977}
Falk, S.W., \& Arnett, W.D. 1977, ApJS, 33, 515

\bibitem[Filippenko(1997)]{filippenko1997}
Filippenko, A.V. 1997, ARAA, 35, 309

\bibitem[Fransson et al.(1996)]{fransson1996}
Fransson, C., Lundqvist, P., \& Chevalier, R.A. 1996, ApJ, 461, 993

\bibitem[Fransson \& Bj\"ornsson(1998)]{fransson1998}
Fransson, C., \& Bj"ornsson, C.-I. 1998, ApJ, 509, 861 

\bibitem[Gezari et al.(2008)]{gezari2008}
Gezari, S., et al. 2008, ApJ, 683, L131

\bibitem[Gr\"afener et al.(2012)]{grafener2012}
Gr\"afener, G., Owocki, S.P., \& Vink, J.S. 2012, A\&A, 538, A40

\bibitem[Hachinger et al.(2011)]{hachinger2011}
Hachinger, S., Mazzali, P.A., Taubenberger, S., Hillebrandt, W., Nomoto, K., Sauer, D.N. 
2011, MNRAS, 422, 70

\bibitem[Horesh et al.(2012)]{horesh2012}
Horesh, A., et al. 2012, preprint (arXiv:1209.1102)

\bibitem[Hunter et al.(2008)]{hunter2008}
Hunter, D.J., et al. 2008, A\&A, 508, 371

\bibitem[Iwamoto et al.(1998)]{iwamoto1998}
Iwamoto, K., et al. 1998, Nature, 395, 672

\bibitem[Krauss et al.(2012)]{krauss2012}
Krauss, M.I., et al. 2012, ApJ, 750, 40

\bibitem[Klein \& Chevalier(1978)]{klein1978}
Klein, R.I., \& Chevalier, R.A. 1978, ApJ, 223, L109

\bibitem[Kulkarni et al.(1998)]{kulkarni1998}
Kulkarni, S.R., et al. 1998, Nature, 395, 663

\bibitem[Lundqvist \& Fransson(1996)]{lundqvist1996}
Lundqvist, P., \& Fransson, C. 1996, ApJ, 464, 924

\bibitem[Maeda et al.(2002)]{maeda2002}
Maeda, K., Nomoto, K., Nakasato, N., \& Suzuki, T. 2002, ASP Conf. Series, 271, 379

\bibitem[Maeda et al.(2006)]{maeda2006}
Maeda, K., Mazzali, P.A., \& Nomoto, K. 2006, ApJ, 645, 1331

\bibitem[Maeda(2012)]{maeda2012}
Maeda, K. 2012, ApJ, in press (arXiv:1209.1466)

\bibitem[Matzner \& McKee(1999)]{matzner1999}
Matzner, C.D., \& McKee, C.F. 1999, ApJ, 510, 379 

\bibitem[Maund et al.(2011)]{maund2011}
Maund, J.R., et al. 2011, ApJ, 739, L37

\bibitem[Mazzali et al.(2002)]{mazzali2002}
Mazzali, P.A., et al. 2002, ApJ, 572, L61

\bibitem[Modjaz et al.(2009)]{modjaz2009}
Modjaz, M., et al. 2009, ApJ, 702, 226

\bibitem[Nomoto et al.(1993)]{nomoto1993}
Nomoto, K., Suzuki, T., Shigeyama, T., Kumagai, S., Yamaoka, H., Saio, H. 1993, 
Nature, 364, 507

\bibitem[Nomoto et al.(2010)]{nomoto2010}
Nomoto, K., Tanaka, M., Tominaga, N., Maeda, K. 2010, New AR, 54, 191

\bibitem[Nugis \& Lamers(2000)]{nugis2000}
Nugis, T., \& Lamers, H.J.G.L.M. 2000, A\&A, 360, 227

\bibitem[Pastorello et al.(2007)]{pastorello2007}
Pastorello, A., et al. 2007, Nature, 447, 829

\bibitem[Rabinak \& Waxman(2011)]{rabinak2011}
Rabinak, I., \& Waxman, E. 2011, ApJ, 728, 63

\bibitem[Sana et al.(2012)]{sana2012}
Sana, H., et al. 2012, Science, 337, 444 

\bibitem[Schavinski et al.(2008)]{schavinski2008}
Schavinski, K., et al. 2008, Science, 321, 223 

\bibitem[Smartt(2009)]{smartt2009}
Smartt, S.J. 2009, ARA\&A, 47, 63

\bibitem[Soderberg et al.(2005)]{soderberg2005}
Soderberg, A.M., Brunthaler, A., Nakar, E., Chevalier, R.A., Bietenholz, M.F. 2005, ApJ, 725, 922

\bibitem[Soderberg et al.(2006a)]{soderberg2006}
Soderberg, A.M., et al. 2006a, Nature, 442, 1014

\bibitem[Soderberg et al.(2006b)]{soderberg2006b}
Soderberg, A.M., Nakar, E., Berger, E., \& Kulkarni, S.R. 2006b, ApJ, 638, 930 

\bibitem[Soderberg et al.(2008)]{soderberg2008}
Soderberg, A.M., et al. 2008, Nature, 453, 469

\bibitem[Soderberg et al.(2010a)]{soderberg2010}
Soderberg, A.M., Brunthaler, A., Naker, E., Chevalier, R.A., \& Bietenholz, M.F. 2010a, ApJ, 725, 922

\bibitem[Soderberg et al.(2010b)]{soderberg2010b}
Soderberg, A.M., et al. 2010b, Nature, 463, 513 

\bibitem[Soderberg et al.(2012)]{soderberg2012}
Soderberg, A.M., et al. 2012, ApJ, 752, 78 

\bibitem[Tominaga et al.(2008)]{tominaga2008}
Tominaga, N., et al. 2008, ApJ, 687, 1208 

\bibitem[Tominaga et al.(2011)]{tominaga2011}
Tominaga, N., et al. 2011, ApJS, 193, 20

\bibitem[Tomita et al.(2006)]{tomita2006}
Tomita, H., et al. 2006, ApJ, 644, 400

\bibitem[Toro(1999)]{toro1999}
Toro, E.F. 1999, Riemann Solvers and Numerical Methods for Fluid Dynamics (Springer)

\bibitem[Valenti et al.(2008)]{valenti2008}
Valenti, S., et al. 2008, ApJ, 673, L155

\bibitem[Van Dyk et al.(2011)]{vandyk2011}
Van Dyk, S.D., et al. 2011, ApJ, 741, L28

\bibitem[Woosley et al.(1999)]{woosley1999}
Woosley, S.E., Eastman, E.G., \& Schmidt, B.P. 1999, ApJ, 516, 788

\bibitem[Woosley \& Bloom(2006)]{woosley2006}
Woosley, S.E., \& Bloom, J.S. 2006, ARA\&A, 44, 507

\bibitem[Yoshii et al.(2003)]{yoshii2003}
Yoshii, Y., et al. 2003, ApJ, 592, 467 

\end{thebibliography}
\end{document}